\documentclass{article}
\usepackage{algorithm}
\usepackage{algorithmic}
\usepackage{amsmath}
\DeclareMathOperator{\sgn}{sgn}
\usepackage{amssymb}
\usepackage{cite}
\usepackage{color}
\usepackage{courier}
\usepackage{graphicx}
\usepackage{mdwlist}
\usepackage{mdwmath}
\usepackage{mdwtab}
\usepackage[caption=false,font=footnotesize]{subfig}
\usepackage{tikz}
\usetikzlibrary{arrows}
\usepackage{times}
\usepackage{url}
\usepackage[normalem]{ulem}

\begin{document}
\title{Efficient Community Detection in Large Networks using Content and Links}
\author{Yiye Ruan, David Fuhry, Srinivasan Parthasarathy \\
Department of Computer Science and Engineering \\
The Ohio State University \\
\{ruan,fuhry,srini\}@cse.ohio-state.edu}
%\numberofauthors{3}
%\author{
%\alignauthor
%Yiye Ruan\\
%\affaddr{Department of Computer Science and Engineering}\\
%\affaddr{The Ohio State University}\\
%\email{ruan@cse.ohio-state.edu}
%\alignauthor
%David Fuhry\\
%\affaddr{Department of Computer Science and Engineering}\\
%\affaddr{The Ohio State University}\\
%\email{fuhry@cse.ohio-state.edu}
%\alignauthor
%Srinivasan Parthasarthy\\
%\affaddr{Department of Computer Science and Engineering}\\
%\affaddr{The Ohio State University}\\
%\email{srini@cse.ohio-state.edu}
%}
\maketitle
%\footnote{Do we need to mention that the work was previously introduced as a poster}
\begin{abstract}
    In this paper we discuss a very simple approach of combining content and link information in graph structures for the purpose of community discovery, a fundamental task in network analysis. Our approach hinges on the basic intuition that many networks contain noise in the link structure and that content information can help strengthen the community signal. This enables ones to eliminate the impact of noise (false positives and false negatives), which is particularly prevalent in online social networks and Web-scale information networks.

    Specifically we introduce a measure of signal strength between two nodes in the network by fusing their link strength with content similarity. Link strength is estimated based on whether the link is likely (with high probability) to reside within a community. Content similarity is estimated through cosine similarity or Jaccard coefficient. We discuss a simple mechanism for fusing content and link similarity.
We then present a biased edge sampling procedure which retains edges that are locally relevant for each graph node. The resulting backbone graph can be clustered using standard community discovery algorithms such as Metis and Markov clustering.

    Through extensive experiments on multiple real-world datasets (Flickr, Wikipedia and CiteSeer) with varying sizes and characteristics, we demonstrate the effectiveness and efficiency of our methods over state-of-the-art learning and mining approaches several of which also attempt to combine link and content analysis for the purposes of community discovery. Specifically we always find a qualitative benefit when combining content with link analysis. Additionally our biased graph sampling approach realizes a quantitative benefit in that it is typically several orders of magnitude faster than competing approaches.
\end{abstract}

%\category{H.2.8}{Database Management}{Database Applications}[Data mining]
\section{Introduction}
An increasing number of applications on the World Wide Web rely on
combining link and content analysis (in different ways) for subsequent 
analysis and inference. For example,
search engines, like Google, Bing and Yahoo! typically use content and 
link information to index,
retrieve and rank web pages. 
Social networking sites like Twitter, Flickr
and Facebook, as well as the aforementioned search engines,
 are increasingly relying on fusing 
content (pictures, tags, text) and link information (friends, followers, 
and users) for deriving actionable knowledge (e.g. marketing and advertising).

In this article we limit our discussion to a fundamental inference problem 
--- that of combining link and content information for the purposes
of inferring clusters or communities of interest. The challenges
are manifold. The topological characteristics of such problems
(graphs induced from the natural link structure) makes identifying 
community structure difficult. Further complicating the issue
is the presence of noise (incorrect links (false positives)
and missing links (false negatives). Determining how to fuse
this link structure with content information efficiently and effectively
is unclear. Finally, underpinning these challenges, is the issue
of scalability as many of these graphs are extremely large running into
millions of nodes and billions of edges, if not larger. 

Given the fundamental nature of this problem, a number of solutions have
emerged in the literature. Broadly these can be classified as: i) those
that ignore content information (a large majority) and focus on addressing
the topological and scalability challenges,
%(exemplars include Metis~\cite{metis}, Graclus~\cite{graclus}, and Markov clustering~\cite{mcl}) 
and ii) those that account for both content and topological information.
%(exemplars include PLSA-PHITS \cite{cohn2001missing}, Link-PLSA-LDA \cite{nallapati2008joint}, Community-User-Topic model \cite{zhou2006probabilistic}, PCL-DC \cite{pcldc}, and SA-Cluster \cite{sacluster})
From a qualitative standpoint the latter presumes to improve on the former
(since the null hypothesis is that content should help improve
the quality of the inferred communities) but 
often at a prohibitive cost to scalability. 

In this article we present CODICIL\footnote{COmmunity Discovery Inferred from Content Information and Link-structure}, a family of highly efficient graph 
simplification algorithms leveraging both content and 
graph topology to identify and retain important edges in a network. 
Our approach relies on fusing content and topological (link) information
in a natural manner.
The output of CODICIL is a
transformed variant of the original graph (with content information),
which can then be clustered by any fast 
content-insensitive graph clustering algorithm such as METIS or Markov clustering.
Through extensive experiments on real-world datasets
drawn from Flickr, Wikipedia, and CiteSeer,
and across several graph clustering algorithms, we demonstrate the effectiveness and efficiency of our methods.
We find that
CODICIL runs several orders of magnitude faster than those state-of-the-art approaches and 
often identifies communities of comparable or superior quality on these datasets.

This paper is arranged as follows. In Section~\ref{sec:relatedwork} we discuss existent research efforts pertaining to our work. The algorithm of CODICIL, along with implementation details, is presented in Section~\ref{sec:methodology}. We report quantitative experiment results in Section~\ref{sec:experiments}, and demonstrate the qualitative benefits brought by CODICIL via case studies in Section~\ref{sec:casestudy}. We finally conclude the paper in Section~\ref{sec:conclusion}.

\section{Related Work}
\label{sec:relatedwork}
\noindent{\bf Community Discovery using Topology (and Content):}
%\subsection{Community Discovery using Topology (and Content)}
Graph clustering/partitioning for community discovery has been studied for more than five decades,
%Graph partitioning techniques Community discovery as well as graph partitioning has already been studied for more than five decades,
and a vast number of algorithms (exemplars include Metis~\cite{metis}, Graclus~\cite{graclus} and Markov clustering~\cite{mcl}) have been proposed and widely used in fields including social network analytics, document clustering, bioinformatics and others. Most of those methods, however, discard content information associated with graph elements. Due to space limitations, we suppress detailed discussions and refer interested readers to recent surveys (e.g. \cite{fortunato2010community}) for a more comprehensive picture. Leskovec et al. compared a multitude of community discovery algorithms based on conductance score, and discovered the trade-off between clustering objective and community compactness~\cite{leskovec2010empirical}. 

Various approaches have been taken to utilize content information for community discovery. One of them is generative probabilistic modeling which considers both contents and links as being dependent on one or more latent variables, and then estimates the conditional distributions to find community assignments. PLSA-PHITS~\cite{cohn2001missing}, Community-User-Topic model \cite{zhou2006probabilistic} and Link-PLSA-LDA~\cite{nallapati2008joint} are three representatives in this category. They mainly focus on studies of citation and email communication networks. Link-PLSA-LDA, for instance, was motivated for finding latent topics in text and citations and assumes different generative processes on citing documents, cited documents as well as citations themselves. Text generation is following the LDA approach, and link creation from a citing document to a cited document is controlled by another topic-specific multinomial distribution. 

Yang et al.~\cite{pcldc} introduced an alternative discriminative probabilistic model, PCL-DC, to incorporate content information in the conditional link model and estimate the community membership directly. In this model, link probability between two nodes is decided by nodes' \emph{popularity} as well as community membership, which is in turn decided by content terms. A two-stage EM algorithm is proposed to optimize community membership probabilities and content weights alternately. Upon convergence, each graph node is assigned to the community with maximum membership probability. 

Researchers have also explored ways to augment the underlying network to take into account the content information. The SA-Cluster-Inc algorithm proposed by Zhou et al.~\cite{saclusterinc}, for example, inserts virtual \textit{attribute nodes} and \textit{attribute edges} into the graph and computes all-pair random walk distances on the new \emph{attribute-augmented graph}. K-means clustering is then used on original graph nodes to assign them to different groups. Weights associated with attributes are updated after each k-means iteration according to their clustering tendencies. The algorithm iterates until convergence.

Ester et al.~\cite{ester2006joint} proposed an heuristic algorithm to solve the \emph{Connected $k$-Center} problem where both connectedness and radius constraints need to be satisfied. The complexity of this method is dependent on the longest distance between any pair of nodes in the feature space, making it susceptible to outliers. Biologists have studied methods \cite{hanisch2002co,ulitsky2007identification} to find functional modules using network topology and gene expression data. Those methods, however, bear domain-specific assumptions on data and are therefore not directly applicable in general.

Recently G{\"u}nnemann et al.~\cite{gunnemann2011db} introduced a subspace clustering algorithm on graphs with feature vectors, which shares some similarity with our topic. Although their method could run on the full feature space, the search space of their algorithm is confined by the intersection, instead of union, of the epsilon-neighborhood and the density-based combined cluster. Furthermore, the construction of both neighborhoods are sensitive to their multiple parameters.

While decent performance can be achieved on small and medium graphs using those methods, it often comes at the cost of model complexity and lack of scalability. Some of them take time proportional to the number of values in each attribute. Others take time and space proportional to the number of clusters to find, which is often unacceptable. Our method, in contrast, is more lightweight and scalable.

\noindent{\bf Clustering/Learning Multiple Graphs:}
%\subsection{Clustering/Learning Multiple Graphs}
Content-aware clustering is also related to multiple-view clustering, as content information and link structure can be treated as two views of the data.
%Clustering from multiple views has been investigated by many researchers.
Strehl and Ghose \cite{strehl2003cluster} discussed three consensus functions (cluster-wise similarity partitioning, hyper-graph partitioning and meta-clustering) to implement cluster ensembles, in which the availability of each individual view's clustering is assumed. Tang et al. \cite{tang2009clustering} proposed a linked matrix factorization method, where each graph's adjacency matrix is decomposed into a ``characteristic'' matrix and a common factor matrix shared among all graphs. The purpose of factorization is to represent each vertex by a lower-dimensional vector and then cluster the vertices using corresponding feature vectors. Their method, while applicable to small-scale problems, is not designed for web-scale networks.
%Multiple views can also be used for semi-supervised learning (also known as \emph{transductive learning}). Detailed discussion is beyond the scope of this paper, and interested readers can refer to representative work including \cite{zhou2007spectral}, \cite{sindhwani2005co}, \cite{argyriou05combining} and \cite{tsuda2005fast}.

\noindent{\bf Graph Sampling for Fast Clustering:}
%\subsection{Graph Sampling for Fast Clustering}
Graph sampling (also known as ``sparsification'' or ``filtering'') has attracted more and more focus in recent years due to the explosive growth of network data. If a graph's structure can be preserved using fewer nodes and/or edges, community discovery algorithms can obtain similar results using less time and memory storage. Maiya and Berger-Wolf~\cite{maiya2010sampling} introduced an algorithm which greedily identifies the node that leads to the greatest \emph{expansion} in each iteration until the user-specified node count is reached. By doing so, an expander-like node-induced subgraph is constructed. After clustering the subgraph, the unsampled nodes can be labeled by using collective inference or other transductive learning methods. This extra post-processing step, however, operates on the original graph as a whole and easily becomes the scalability bottleneck on larger networks.

Satuluri et al.~\cite{lspar} proposed an edge sampling method to preferentially retain edges that connect two similar nodes. The localized strategy ensures that edges in the relatively sparse areas will not be over-pruned. Their method, however, does not consider content information either.

Edge sampling has also been applied to other graph tasks. Karger \cite{karger1999random} studied the impact of random edge sampling on original graph's cuts, and proposed randomized algorithms to find graph's minimum cut and maximum flow. Aggarwal et al. \cite{aggarwal2011outlier} proposed using edging sampling to maintain structural properties and detect outliers in graph streams. The goals of those work are not to preserve community structure in graphs, though.

\section{Methodology}
\label{sec:methodology}
We begin by defining the notations used in the rest of our paper.
Let $\mathcal{G}_t = (\mathcal{V}, \mathcal{E}_t, \mathcal{T})$ be an undirected graph with $n$ vertices $\mathcal{V} = v_1, \ldots, v_n$, edges $\mathcal{E}_t$, and a collection of $n$ corresponding term vectors $\mathcal{T} = \boldsymbol{t}_1, \ldots, \boldsymbol{t}_n$.
We use the terms ``graph'' and ``network'' interchangeably as well as the terms ``vertex'' and ``node''.
%Note that ``graph'' and ``network'' will be used exchangeably throughout the paper if no confusion was caused, and so are ``vertex'' and ``node''. 
Elements in each term vector $\boldsymbol{t}_i$ are basic content units which can be single words, tags or \textit{n}-grams, etc., depending on the context of underlying network. 
For each graph node $v_i\in\mathcal{V}$, let its term vector be $\boldsymbol{t}_i$.
%\subseteq \mathcal{C}$. \\

Our goal is to generate a simplified, edge-sampled graph $\mathcal{G}_{sample} = (\mathcal{V}, \mathcal{E}_{sample})$ and then use $\mathcal{G}_{sample}$ to find communities with coherent content and link structure. $\mathcal{G}_{sample}$ should possess the following properties:
\begin{itemize*}
    \item
        $\mathcal{G}_{sample}$ has the same vertex set as $\mathcal{G}_t$. That is, no node in the network is added or removed during the simplification process.
    \item
        $|\mathcal{E}_{sample}| \ll |\mathcal{E}_t|$, as this enables both better runtime performance and lower memory usage in the subsequent clustering stage.
    \item
        Informally put, the resultant edge set $\mathcal{E}_{sample}$ would connect node pairs which are both structure-wise and content-wise similar. As a result, it is possible for our method to add edges which were absent from $\mathcal{E}_t$ since the content similarity was overlooked.
\end{itemize*}
%which shares the same vertex set with $\mathcal{G}_t$. The edge set $\mathcal{E}_s$ is not necessarily a subset of $\mathcal{E}_t$, since CANS is likely to insert edges connecting highly content-wise similar node pairs. However, it is required that $|\mathcal{E}_s| \ll |\mathcal{E}_t|$ as doing so enables significantly lower temporal-spatial cost in the sequent clustering stage. \\

\subsection{Key Intuitions}
The main steps of the CODICIL algorithm are:
\begin{enumerate*}
    \item Create content edges.
    \item Sample the union of content edges and topological edges with bias, retaining only edges that are relevant in local neighborhoods.
    \item Partition the simplified graph into clusters.
\end{enumerate*}

\begin{figure*}[thb]
    \centering
    \tikzstyle{int}=[draw, fill=blue!20, minimum size=2em]
%\tikzstyle{init} = [pin edge={to-,thin,black}]

    \begin{tikzpicture}[node distance=2.5cm,auto,>=latex']
%, pin={[init]above:$v_0$}
        \node[int] at (-1,1.75) (terms) {Term vectors $\mathcal{T}$};
        \node[int] at (2,0.25) (topo) {Topological edges $\mathcal{E}_t$};
        \node[int] at (2,1.75) (cont)  {Content edges $\mathcal{E}_{c}$};
        \node[int] at (4.75,1) (union) {Edge union $\mathcal{E}_{u}$};
%, pin={[init]below:cluster}
        \node[int] at (7.75,1) (sampl)  {Edge subset $\mathcal{E}_{sample}$};
        \node[int] at (7.5,-0.5) (nodes)  {Vertices $\mathcal{V}$};
        \node[int] at (10,0.25) (clusters) {Clustering $\mathcal{C}$};
        \node at (0.25,2.5) {1. Create content edges};
        \node at (2.25,1) {2. Combine edges};
        \node at (6,1.75) {3. Sample edges with bias};
        \node at (8.25,0.25) {4. Cluster};
        %\node[int] at (3.5,6) (terms) {Term vectors $\mathcal{T}$};
        %\node[int] at (0,4) (topo) {Topological edges $\mathcal{E}_t$};
        %\node[int] at (3.5, 4) (cont)  {Content edges $\mathcal{E}_{c}$};
%%, pin={[init]below:cluster}
        %\node[int] at (1.5,2) (sampl)  {Sampled edges $\mathcal{E}_{sample}$};
        %\node[int] at (1.5,0) (nodes)  {Vertices $\mathcal{V}$};
        %\node[int] at (6, 1) (clusters) {Clustering $\mathcal{C}$};
        %\node at (1.5,3) {2. Combine \& sample edges};
        %\node at (4,1) {3. Cluster};
    %\node at (3.5,5) {1. Create content edges};
    %\node (b) [above of=a,node distance=2cm, coordinate] {a};
%, pin={[init]above:$p_0$}
    %\node [coordinate] (end) [right of=c, node distance=2cm]{};
    %\path[->] (b) edge node {$a$} (a);
        \path[->] (terms) edge node {} (cont);
        \path[->] (topo) edge node {} (union);
        \path[->] (cont) edge node {} (union);
        \path[->] (union) edge node {} (sampl);
        \path[->] (sampl) edge node {} (clusters);
        \path[->] (nodes) edge node {} (clusters);
    %\draw[->] (c) edge node {$p$} (end) ;
    \end{tikzpicture}
    \caption{Work flow of CODICIL}
    \label{fig:edgeflow}
\end{figure*}
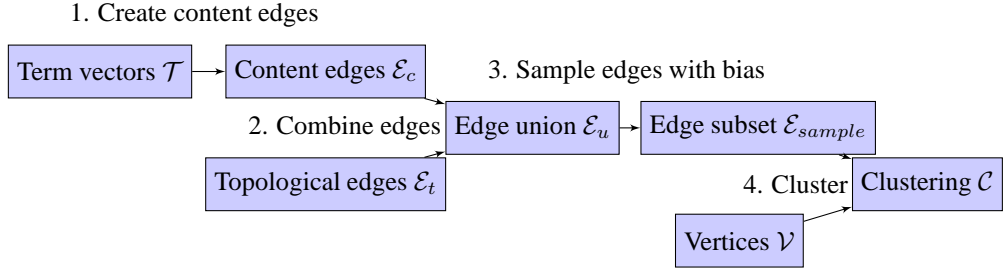

The constructed content graph and simplified graph have the same vertices as the input graph (vertices are never added or removed), so the essential operations of the algorithm are constructing, combining edges and then sampling with bias. Figure~\ref{fig:edgeflow} illustrates the work flow of CODICIL.

%and produces a partitioning of vertices in three main steps.
%three main steps to produce a partitioning of vertices from a content-annotated input graph.
%Given a content-annotated input graph and parameters, the CODICIL algorithm produces a partitioning of vertices into clusters in three steps.
From the term vectors $\mathcal{T}$, content edges $\mathcal{E}_c$ are constructed.
Those content edges and the input topological edges $\mathcal{E}_t$ are combined as $\mathcal{E}_u$ which is then sampled with bias to form a smaller edge set $\mathcal{E}_{sample}$ where the most relevant edges are preserved.
The graph composed of these sampled edges is passed to the graph clustering algorithm which partitions the vertices into a given number of clusters.
%In first step of CODICIL, a content graph $\mathcal{G}_c$ is constructed. The top-k neighbors of each vertex are found and add edges to them are added to the content edge set $\mathcal{E}_c$.
%Secondly, an edge-sampled graph $\mathcal{G}_{sampl}$ is constructed from a sampling of the combination of the newly constructed content edges $\mathcal{E}_c$ and the toplological edge set $\mathcal{E}_t$.
%For each vertex $v_i$ its relationship weight with each of its neighbors $v_j \in ngh(v_i)$ is computed by taking the Jaccard coefficient of the neighbor sets of $v_i$ and each $v_j$, separately for the topological and content graphs.
%The weight distributions for each graph are normalized, then combined and ranked, and the edges with largest combined weight are preserved and added to the sampled edge set $\mathcal{E}_{sampl}$.
%Finally, the sampled graph $\mathcal{G}_{sampl}$ is passed to the graph clustering algorithm which partitions the vertices into (given) $l$ clusters.
%namely, to in step 1 create, and in step 2 combine \& filter edges. Step 3, the partitioning, is performed by a given graph clustering algorithm.
\subsection{Basic Framework}
The pseudo-code of CODICIL is given in Algorithm~\ref{alg:codicil}.
\mathchardef\mhyphen="2D
\begin{algorithm}[ht!]
\caption{CODICIL}
\label{alg:codicil}
\begin{algorithmic}[1]
\REQUIRE $\mathcal{G}_t=(\mathcal{V},\mathcal{E}_t,\mathcal{T})$, $k$, $normalize(\cdot)$, $\alpha \in [0,1]$, $l$, $clusteralgo(\cdot, \cdot)$, $similarity(\cdot, \cdot)$
\RETURNS $\mathcal{C}$ (a disjoint clustering of $\mathcal{V})$
\STATE \COMMENT{Create content edges $\mathcal{E}_c$}
\STATE $\mathcal{E}_c \leftarrow \emptyset$
%\FORALL{$v \in \mathcal{V}$}
\FOR{$i=1$ to $|\mathcal{V}|$}
\FORALL{$v_j \in TopK(v_i, k, \mathcal{T})$}
\STATE $\mathcal{E}_c \leftarrow \mathcal{E}_c \cup (v_i, v_j)$
\ENDFOR
\ENDFOR
\STATE \COMMENT{Combine $\mathcal{E}_t$ and $\mathcal{E}_c$. Retain edges with a bias towards locally relevant ones}
%\STATE \COMMENT{Combine $\mathcal{G}_t$ and $\mathcal{G}_c$. Construct the sampled graph $\mathcal{G}_{sample}$}
\STATE $\mathcal{E}_u \leftarrow \mathcal{E}_t \cup \mathcal{E}_c$
%\STATE $\mathcal{G}_c \leftarrow (\mathcal{V}, \mathcal{E}_c)$
\STATE $\mathcal{E}_{sample} \leftarrow \emptyset$
%\FORALL{$v \in \mathcal{V}$}
\FOR{$i=1$ to $|\mathcal{V}|$}
\STATE \COMMENT{$\Gamma_i$ contains $v_i$'s neighbors in the edge union}
%\STATE \COMMENT{$\Gamma_i$ is the union of $v_i$'s topology graph neighbors and content graph neighbors}
\STATE $\Gamma_i \leftarrow ngbr(v_i, \mathcal{E}_u)$
%\STATE $\Gamma_i \leftarrow ngbr(v_i, \mathcal{E}_t) \cup ngbr(v_i, \mathcal{E}_c)$
\STATE \textbf{for} $j=1$ to $|\Gamma_i|$ \textbf{do} $\boldsymbol{sim^t}_{ij} \leftarrow similarity(ngbr(v_i, \mathcal{E}_t), ngbr({\gamma}_j, \mathcal{E}_t))$
%\frac{\# matches}{\# hashes}$
\STATE $\boldsymbol{simnorm^t}_i \leftarrow normalize(\boldsymbol{sim^t}_i)$
\STATE \textbf{for} $j=1$ to $|\Gamma_i|$ \textbf{do} $\boldsymbol{sim^c}_{ij} \leftarrow similarity(\boldsymbol{t}_i, \boldsymbol{t}_{{\gamma}_j})$
\STATE $\boldsymbol{simnorm^c}_i \leftarrow normalize(\boldsymbol{sim^c}_i)$
\STATE \textbf{for} $j=1$ to $|\Gamma_i|$ \textbf{do} $\boldsymbol{sim}_{ij} \leftarrow \alpha \cdot \boldsymbol{simnorm^t}_{ij} + (1-\alpha) \cdot \boldsymbol{simnorm^c}_{ij}$
\STATE \COMMENT{Sort similarity values in descending order. Store the corresponding node IDs in $idx_i$}
\STATE $[val_i, idx_i] \leftarrow descsort(\boldsymbol{sim}_i)$ 
\FOR{$j=1$ to $\left \lceil \sqrt{|\Gamma_i|} \right \rceil$}
\STATE $\mathcal{E}_{sample} \leftarrow \mathcal{E}_{sample} \cup (v_i, v_{idx_{ij}})$
\ENDFOR
\ENDFOR
\STATE $\mathcal{G}_{sample} \leftarrow (\mathcal{V}, \mathcal{E}_{sample})$
\STATE $\mathcal{C} \leftarrow clusteralgo(\mathcal{G}_{sample}, l)$ \COMMENT{Partition into $l$ clusters}
\RETURN $\mathcal{C}$
\end{algorithmic}
\end{algorithm}

CODICIL takes as input 1) $\mathcal{G}_t$, the original graph consisting of vertices $V$, edges $\mathcal{E}_t$ and term vectors $\mathcal{T}$ where $\boldsymbol{t}_i$ is the content term vector for vertex $v_i$, $1 \leq i \leq |\mathcal{V}| = |\mathcal{T}|$, 2) $k$, the number of nearest content neighbors to find for each vertex, 3) $normalize(\boldsymbol{x})$, a function that normalizes a vector $\boldsymbol{x}$, 4) $\alpha$, an optional parameter that specifies
the weights of topology and content similarities, 5) $l$, the number of output clusters desired, 6) $clusteralgo(\mathcal{G}, l)$, an algorithm that partitions a graph $\mathcal{G}$ into $l$ clusters and 7) $similarity(\boldsymbol{x}, \boldsymbol{y})$ to compute similarity between $\boldsymbol{x}$ and $\boldsymbol{y}$. Note that any content-insensitive graph clustering algorithm can be plugged in the CODICIL framework, providing great flexibility for applications. 
\subsubsection{Creating Content Edges}
\label{sec:k}
Lines 2 through 7 detail how content edges are created.
%First a content graph $\mathcal{G}_c$ is built (lines 2 through 8). 
For each vertex $v_i$, its $k$ most content-similar neighbors are computed\footnote{Besides top-$k$ criteria, we also investigated using all-pairs similarity above a given global threshold, but this tended to produce highly imbalanced degree distributions.}. For each of $v_i$'s top-$k$ neighbors $v_j$, an edge $(v_i, v_j)$ is added to content edges $\mathcal{E}_c$. 
%\textcolor{red}{Maybe we should redescribe ``knn'' as ``top-k'', since in practice we approximated knn vectors with top tf/idf terms which gave better results, and is more similar to text search top-k ranking algorithms.}
In our experiments we implemented the $TopK$ sub-routine by calculating the cosine similarity of $\boldsymbol{t}_i$'s TF-IDF vector and each other term vector's TF-IDF vector. For a content unit $c$, its TF-IDF value in a term vector $\boldsymbol{t}_i$ is computed as
\begin{equation}
    tf \mhyphen idf(c,\boldsymbol{t}_i) = \sqrt{tf(c,\boldsymbol{t}_i)} \cdot \log{\left( 1+\frac{|\mathcal{T}|}{\sum_{j=1}^{|\mathcal{T}|}{tf(c,\boldsymbol{t}_j)}}\right)} \enspace .
\end{equation}

The cosine similarity of two vectors $\boldsymbol{x}$ and $\boldsymbol{y}$ is
\begin{equation}
    \label{eq:CosSim}
    cosine(\boldsymbol{x}, \boldsymbol{y}) = \frac{\boldsymbol{x} \cdot \boldsymbol{y}}{\|\boldsymbol{x}\|_2 \cdot \|\boldsymbol{y}\|_2} \enspace .
\end{equation}
%The cosine similarity of two TF-IDF vectors $\boldsymbol{t'}_i$ and $\boldsymbol{t'}_j$ is
%\begin{equation}
    %\label{eq:CosSim}
    %cos \mhyphen sim(\boldsymbol{t'}_i, \boldsymbol{t'}_j) = \frac{\boldsymbol{t'}_i \cdot \boldsymbol{t'}_j}{\|\boldsymbol{t'}_i\|_2 \cdot \|\boldsymbol{t'}_j\|_2} \enspace .
%\end{equation}

The $k$ vertices corresponding to the $k$ highest TF-IDF vector cosine similarity values with $v_i$ are selected as the top-$k$ neighbors of $v_i$.
\subsubsection{Local Ranking of Edges and Graph Simplification}
Line 9 takes the union of the newly-created content edge set $\mathcal{E}_c$ and the original topological edge set $\mathcal{E}_t$. In lines 10 through 24, a sampled edge set $\mathcal{E}_{sample}$ is constructed by retaining the most relevant edges from the edge union $\mathcal{E}_u$. For each vertex $v_i$, the edges to retain are selected from its local neighborhood in $\mathcal{E}_u$ (line 13). We compute the topological similarity (line 14) between node $v_i$ and its neighbor $\gamma_j$ as the relative overlap of their respective topological neighbor sets, $I = ngbr(v_i, \mathcal{E}_t)$ and $J = ngbr(\gamma_j, \mathcal{E}_t)$, using $similarity$ (either cosine similarity as in Equation \ref{eq:CosSim} or Jaccard coefficient as defined below):
\begin{equation}
\label{eq:jaccard}
jaccard(I, J) = \frac{|I \cap J|}{|I \cup J|} \enspace .
\end{equation}

After the computation of the topological similarity vector $\boldsymbol{sim^t}_i$ finishes, it is normalized by $normalize$ (line 15). 
In our experiments we implemented $normalize$ with either $zero \mhyphen one$, which simply rescales the vector to $[0,1]$:
\begin{equation}
    zero \mhyphen one(\vec{x}) =  (x_i - min(\vec{x})) / (max(\vec{x}) - min(\vec{x}))
\end{equation}
or $z \mhyphen norm$\footnote{Montague and Aslam \cite{montague2001relevance} pointed out that $z \mhyphen norm$ has the advantage of being both shift and scale invariant as well as outlier insensitive. They experimentally found it best among six simple combination schemes discussed in~\cite{fox1994combination}.}, which centers and normalizes values to zero mean and unit variance:
\begin{equation}
    z \mhyphen norm(\vec{x}) = \frac{x_i - \hat{\mu}}{\hat{\sigma}}, \hat{\mu} = \frac{\sum_{i=1}^{|\vec{x}|} x_i}{|\vec{x}|}, \hat{\sigma}^2 = \frac{1}{|\vec{x}|-1}\sum_{i=1}^{|\vec{x}|} (x_i - \hat{\mu})^2 \enspace .
\end{equation}
Likewise, we compute $v_i$'s content similarity to its neighbor ${\gamma}_j$ by applying $similarity$ on term vectors $\boldsymbol{t}_i$ and $\boldsymbol{t}_{\gamma_j}$ and normalize those similarities (lines 16 and 17).
The topological and content similarities of each edge are then aggregated with the weight specified by $\alpha$ (line 18).
%For \emph{z-norm}, the topological and content similarities of each edge are combined by default with equal weights (line 17). The parameterized variant of the algorithm assigns a weight of $\alpha$ to topology and $(1-\alpha)$ to content.

In lines 20 through 23, the edges with highest similarity values are retained.
As stated in our desiderata, we want $|\mathcal{E}_{sample}| \ll |\mathcal{E}_t|$ and therefore need to retain fewer than $|\Gamma_i|$ edges. Inspired by \cite{lspar}, we choose to keep $\lceil \sqrt{|\Gamma_i|} \rceil$ edges. This form has the following properties: 
1) every vertex $v_i$ will be incident to at least one edge, therefore the sparsification process does not generate new singleton,
%1) every $v_i$ will be incident to at least one edge and never become disconnected, 
2) concavity and monotonicity ensure that larger-degree vertices will retain no fewer edges than smaller-degree vertices, and 3) sublinearity ensures that smaller-degree vertices will have a larger fraction of their edges retained than larger-degree vertices.
%We choose to retain square root of the degree of $v_i$ many edges () for the following reasons: 1) square root many edges because 1) every vertex will retain at least one incident edge, 2) it is concave and monotonically nondecreasing so vertices with larger degree retain no fewer samples than vertices with smaller degree, 3) it is sublinear so vertices with larger degree retain a lower fraction of edges than vertices with smaller degree
\subsubsection{Partitioning the Sampled Graph}
Finally in lines 25 through 27 the sampled graph $\mathcal{G}_{sample}$ is formed with the retained edges, and the graph clustering algorithm $clusteralgo$ partitions $\mathcal{G}_{sample}$ into $l$ clusters.

\subsubsection{Extension to Support Complex Graphs}
The proposed CODICIL framework can also be easily extended to support community detection from other types of graph. If an input graph has weighted edges, we can modify the formula in line 18 so that $sim_{ij}$ becomes the product of combined similarity and original edge weight. Support of attribute graph is also straightforward, as attribute assignment of a node can be represented by an indicator vector, which is in the same form of a text vector.

\subsection{Key Speedup Optimizations}
\subsubsection{\ensuremath{\boldsymbol{TopK}} Implementation}
\label{sec:m}
When computing cosine similarities across term vectors $\boldsymbol{t}_1, \ldots, \boldsymbol{t}_{|\mathcal{T}|}$, one can truncate the TF-IDF vectors by only keeping $m$ elements with the highest TF-IDF values and set other elements to 0. When $m$ is set to a small value, TF-IDF vectors are sparser and therefore the similarity calculation becomes more efficient with little loss in accuracy.

We may also be interested in constraining content edges to be within a topological neighborhood of each node $v_i$, such that the search space of $TopK$ algorithm can be greatly reduced. Two straightforward choices are 1) ``1-hop'' graph in which the content edges from $v_i$ are restricted to be in $v_i$'s direct topological neighborhood, and 2) ``2-hop'' graph in which content edges can connect $v_i$ and its neighbors' neighbors.  

Many contemporary text search systems make use of inverted indices to speed up the operation of finding the $k$ term vectors (documents) with the largest values of Equation \ref{eq:CosSim} given a query vector $\boldsymbol{t}_i$. We used the implementation from Apache Lucene for the largest dataset. 
%(\ref{eq:CosSim}) and selection of the $k$ documents (term vectors) with the largest $CosSim_i(d)$ values.
%Rather than considering each document $j$ in the corpus, contemporary text search tools use inverted indices to perform the top-k 
%Contemporary text search tools use inverted indices to speed up the 
%This process is sped up in contemporary document indexing systems such as Lucene, 
\subsubsection{Fast Jaccard Similarity Estimation}
\label{subsec:jaccard_est}
To avoid expensive computation of the exact Jaccard similarity, we estimate it by using minwise hashing~\cite{broder1998min}. An unbiased estimator of sets $A$ and $B$'s Jaccard similarity can be obtained by
\begin{equation}
    \hat{jaccard}(A,B) = \frac{1}{h}\sum_{i=1}^{h}{I(\min(\pi_{i}(A))=\min(\pi_{i}(B)))} \enspace ,
    \label{eq:minwise}
\end{equation}
where $\pi_1, \pi_2,\cdots,\pi_h$ are $h$ permutations drawn randomly from a family of minwise independent permutations defined on the universe $A$ and $B$ belong to, and $I$ is the identity function. After hashing each element once using each permutation, the cost for similarity estimation is only $O(h)$ where $h$ is usually chosen to be less than $|A|$ and $|B|$. 

\subsubsection{Fast Cosine Similarity Estimation}
Similar to Jaccard coefficient, we can apply random projection method for fast estimate of cosine similarity \cite{charikar2002similarity}. In this method, each hash signature for a $d$-dimensional vector $\boldsymbol{x}$ is $h(\boldsymbol{x}) = \sgn{(\boldsymbol{x}, \boldsymbol{r})}$, where $\boldsymbol{r} \in \{0,1\}^d$ is drawn randomly. For two vectors $\boldsymbol{x}$ and $\boldsymbol{y}$, the following holds:
\begin{equation}
    Pr[h(\boldsymbol{x}) = h(\boldsymbol{y})] = 1 - \frac{\arccos{(cosine(\boldsymbol{x}, \boldsymbol{y}))}}{\pi} \enspace .
\end{equation}
\subsection{Performance Analysis}
Lines 3--7 of CODICIL are a preprocessing step which compute for each vertex its top-$k$ most similar vertices. Results of this one-time computation can be reused for any $k^\prime \leq k$. Its complexity depends on the implementation of the $TopK$ operation. On our largest dataset Wikipedia this step completed within a few hours.

We now consider the loop in lines 11--24 where CODICIL loops through each vertex. 
For lines 14 and 16 we use the Jaccard estimator from Section~\ref{subsec:jaccard_est} for which runs in $O(h)$ with a constant number of hashes $h$.
The normalizations in lines 15 and 17 are $O(|\Gamma_i|)$ and the inner loop in lines 21--23 is $O(\sqrt{|\Gamma_i|})$. Sorting edges by weight in line 20 is $O(|\Gamma_i| \log{|\Gamma_i|})$. The size of $\Gamma_i$, the union of topology and content neighbors, is at most $n$ but on average much smaller in real world graphs. Thus the loop in lines 11--24 runs in $O(n^2 \log{n})$.
%(see column ``Avg $|\boldsymbol{t}_i|$'' of Table~\ref{tab:dataset})

The overall runtime of CODICIL is the edge preprocessing time, plus $O(n^2 \log{n})$ for the loop, plus the algorithm-dependent time taken by $clusteralgo$.

%\subsection{Remarks}
%We conclude this section with two observations.
%Biased edge sampling removes spurious connections between nodes and speeds up subsequent computation.
%If we view a normalized adjacency matrix as a random walk transition matrix, the intuition of clustering is to find groups of which a random surfer rarely ``walks out''. By removing spurious edges with low combined similarity and making inter-cluster walks more difficult, CODICIL attempts to discover more cohesive communities. To the best of our knowledge, this is the first application of biased edge sampling to content-aware graph clustering.

%Previous work has shown that simply adding up the adjacency matrices of $\mathcal{G}_t$ and $\mathcal{G}_c$ yields sub-par performance \cite{zhou2007spectral,tang2009clustering}. By contrast, in CODICIL each edge's similarity always comprises both structural and semantic similarity values, because similarity fusion is performed in each node's local neighborhood in the union graph $(\mathcal{V}, \mathcal{E}_u)$ (see line 13 of Algorithm \ref{alg:codicil}).
%First of all, similarity fusion in CODICIL is more than simply adding up the adjacency matrices of $\mathcal{G}_t$ and $\mathcal{G}_c$, which has been shown to have sub-par performance \cite{zhou2007spectral,tang2009clustering}.

\section{Experiments}
\label{sec:experiments}
We are interested in empirically answering the following questions:
\begin{itemize*}
    \item
        {\bf Do the proposed content-aware clustering methods lead to better clustering than using graph topology only?}
    \item
        {\bf How do our methods compare to existing content-aware clustering methods?}
    \item
        {\bf How scalable are our methods when the data size grows?}
\end{itemize*}
\subsection{Datasets}
Three publicly-available datasets with varying scale and characteristic are used. Their domains cover document network as well as social network. Each dataset is described below, and Table \ref{tab:dataset} follows, listing basic statistics of them. 
\subsubsection{CiteSeer}
A citation network of computer science publications\footnote{\url{http://www.cs.umd.edu/projects/linqs/projects/lbc/index.html}}, each of which labeled as one of six sub-fields. In our graph, nodes stand for publications and undirected edges indicate citation relationships. The content information is stemmed words from research papers, represented as one binary vector for each document. Observe that the density of this network (average degree 2.74) is significantly lower than normally expected for a citation network.
\subsubsection{Wikipedia}
        %\textbf{Wikipedia:} The October 2011 dump of English Wikipedia pages\footnote{\url{http://en.wikipedia.org/wiki/Wikipedia:Database_download#English-language_Wikipedia}}. Only regular pages which belong to at least one category are kept, and each of which is represented as one vertex in the graph. Inter-page links are extracted and treated as undirected. Cleaned bigrams from title and text are used to represent each document's content. 
The static dump of English Wikipedia pages (October 2011). Only regular pages belonging to at least one category are included, each of which becomes one node. Page links are extracted. Cleaned bi-grams from title and text are used to represent each document's content. We use categories that a page belongs to as the page's class labels. Note that a page can be contained in more than one category, thus ground truth categories are overlapping.% In Table \ref{tab:dataset} we show both the total number of ground truth groups as well as the number of groups that has at least one hundred pages.}
\subsubsection{Flickr}
From a dataset of tagged photos\footnote{\url{http://staff.science.uva.nl/~xirong/index.php?n=DataSet.Flickr3m}} we removed infrequent tags and users associated with only few tags. Each graph node stands for a user, and an edge exists if one user is in another's contact list. Tags that users added to uploaded photos are used as content information. Flickr user groups are collected as ground truth. Similar to Wikipedia categories, Flickr user groups are also overlapping.
        %and consider users who have used more than 50 unique tags in their photos. Each user is modeled as one node in the graph. Flickr allows users to add others into a \emph{contact list}\footnote{\url{http://www.flickr.com/help/contacts/}} to share photos, which can be viewed as an indication of the social relationship among users and add an undirected edge between two nodes if one user is in another's contact list. Among those user-generated tags, only the ones which are used by more than 200 users are retained. Flickr also allows users to join a vast selection of interest groups. We used Flickr API to collect the list of 184,421 groups for which at least one user in our dataset is a member of.
%{\bf Please reduce the description for Flickr}
\begin{table*}[ht]
    %\centering
    \hspace{-0.5in}
    \begin{tabular}{|c|c|c|c|c|c|c|c|}
        \hline
        & $|\mathcal{V}|$ & $|\mathcal{E}_t|$ & \# CC & $|\mbox{CC}_{\mbox{max}}|$ & \# Uniq. Content Unit & Avg $|\boldsymbol{t}_i|$ & \# Class \\ \hline
        Wikipedia & 3,580,013 & 162,085,383 & 10 & 3,579,995 & 1,459,335 & 202 & 595,355 \\ \hline
        Flickr & 16,710 & 716,063 & 4 & 16,704 & 1,156 & 44 & 184,334 \\ \hline
        CiteSeer & 3,312 & 4,536 & 438 & 2,110 & 3,703 & 32 & 6 \\ \hline
    \end{tabular}
    %\begin{tabular}{|c|c|c|c|c|c|c|c|}
        %\hline
        %& $|\mathcal{V}|$ & $|\mathcal{E}_t|$ & $\# CC$ & $|CC_{max}|$ & Clustering Coefficient & $|\mathcal{T}|$ & Avg $|t_i|$ \\ \hline
        %Wikipedia & 3,580,013 & 162,085,383 & 10 & 3,579,995 & & 1,459,335 & 202 \\ \hline
        %Flickr & 16,710 & 716,063 & 4 & 16,704 & 0.1070 & 1,156 & 44 \\ \hline
        %CiteSeer & 3,312 & 4,536 & 438 & 2,110 & 0.1426 & 3,703 & 32 \\ \hline
    %\end{tabular}
    \caption{Basic statistics of datasets. \# CC: number of connected components. \ensuremath{|\mbox{CC}_{\mbox{max}}|}: size of the largest connected component. Avg \ensuremath{|\boldsymbol{t}_i|}: average number of non-zero elements in term vectors. \# Class: number of (overlapping) ground truth classes.}
    \label{tab:dataset}
\end{table*}
\subsection{Baseline Methods}
In terms of strawman methods, we compare the CODICIL methods with three existing content-aware graph clustering algorithms, SA-Cluster-Inc \cite{saclusterinc}, PCL-DC \cite{pcldc} and Link-PLSA-LDA (L-P-LDA) \cite{nallapati2008joint}. Their methodologies have been briefly introduced in Section~\ref{sec:relatedwork}. When applying SA-Cluster-Inc, we treat each term in $\mathcal{T}$ as a binary-valued attribute, i.e. for each graph node $i$ every attribute value indicates whether the corresponding term is present in $\boldsymbol{t}_i$ or not. For L-P-LDA, since it does not assume a distinct distribution over topics for each cited document individually, only citing documents' topic distributions are estimated. As a result, there are 2313 citing documents in CiteSeer dataset and we report the F-score on those documents using their corresponding ground-truth assignments.

Previously SA-Cluster-Inc has been shown to outperform k-SNAP \cite{Tian2008} and PCL-DC to outperform methods including PLSA-PHITS \cite{cohn2001missing}, LDA-Link-Word \cite{erosheva2004mixed} and Link-Content-Factorization \cite{zhu2007combining}. Therefore we do not compare with those algorithms.
%We obtained both source code packages from the respective authors.

Two content-insensitive clustering algorithms are included in the experiments as well. The first method, ``Original Topo'', clusters the original network directly. The second method samples edges solely based on structural similarity and then clusters the sampled graph \cite{lspar}, and we refer to it as ``Sampled Topo'' hereafter.

Finally, we also adapt LDA and K-means\footnote{We do not report running time of K-means as it is not implemented in C or C++.} algorithm to cluster graph nodes using content information only. When applying LDA, we treat each term vector $\boldsymbol{t}_i$ as a document, and one product of LDA's estimation procedure is the distribution over latent topics, $\boldsymbol{\theta}_{\boldsymbol{t}_i}$, for each $\boldsymbol{t}_i$ (more details can be found at the original paper by Blei et al.~\cite{lda}). Therefore, we treat each latent topic as a cluster and assign each graph node to the cluster that corresponds to the topic of largest probability. We use GibbsLDA++\footnote{\url{http://gibbslda.sourceforge.net/}}, a C++ implementation of LDA using Gibbs sampling~\cite{gibbs} which is faster than the variational method proposed originally. Results of this method are denoted as ``LDA''.
\subsection{Experiment Setup}
\subsubsection{Parameter Selection}
There are several tunable parameters in the CODICIL framework, first of which is $k$, the number of content neighbors in the $TopK$ sub-routine. We propose the following heuristic to decide a proper value for $k$: the value of $k$ should let $|\mathcal{E}_c| \approx |\mathcal{E}_t|$. As a result, $k$ is set to 50 for both Wikipedia ($|\mathcal{E}_c| = 150,955,014$) and Flickr ($|\mathcal{E}_c| = 722,928$). For CiteSeer, we experiment with two relatively higher $k$ values (50, $|\mathcal{E}_c| = 103,080$ and 70, $|\mathcal{E}_c| = 143,575$) in order to compensate the extreme sparsity in the original network. Though simplistic, this heuristic leads to decent clustering quality, as shown in Section \ref{subsec:clustering_quality}, and avoids extra effort for tuning.

Another parameter of interest is $\alpha$, which determines the weights for structural and content similarities. We set $\alpha$ to 0.5 unless otherwise specified, as in Section \ref{sec:vary_alpha}. The number of hashes ($h$) used for minwise hashing (Jaccard coefficient) is 30, and 512 for random projection (cosine similarity). Experiments with both choices of $similarity$ function are performed. As for $m$, the number of non-zero elements in term vectors, we let $m=10$ for Wikipedia and Flickr. This optional step is omitted for CiteSeer since the speedup is insignificant.
%We use two different strategies in experiments:
%\begin{itemize*}
    %\item Normalize similarity values using \emph{z-norm} and fix $\alpha$ at 0.5 (i.e. give two parts equal weights). We denote this strategy as ``zNorm''.
    %\item Normalize similarity values using \emph{zero-one} and vary $\alpha$ from 0 to 1 with a step length 0.05. We will report the best quality measure over $\alpha$'s value, and referred to it as ``Max''.
%\end{itemize*}

%As for $m$, the number of non-zero elements in term vectors, we let $m=10$ for Wikipedia and Flickr. This optional step is omitted for CiteSeer since the speedup is insignificant. The number of hashes used for minwise hashing ($h$) is fixed at 30.
\subsubsection{Clustering Algorithm}
We combine the CODICIL framework with two different clustering algorithms, Metis\footnote{\url{http://glaros.dtc.umn.edu/gkhome/metis/metis/download}}~\cite{metis} and Multi-level Regularized Markov Clustering (MLR-MCL)\footnote{\url{http://www.cse.ohio-state.edu/~satuluri/research.html}}~\cite{mlrmcl}. Both clustering algorithms are also applied on strawman methods.
\subsection{Effect of Simplification on Graph Structure}
\begin{figure}[htb]
    \centering
    \subfloat[Citeseer]
    {
        \includegraphics[angle=-90,width=0.45\textwidth]{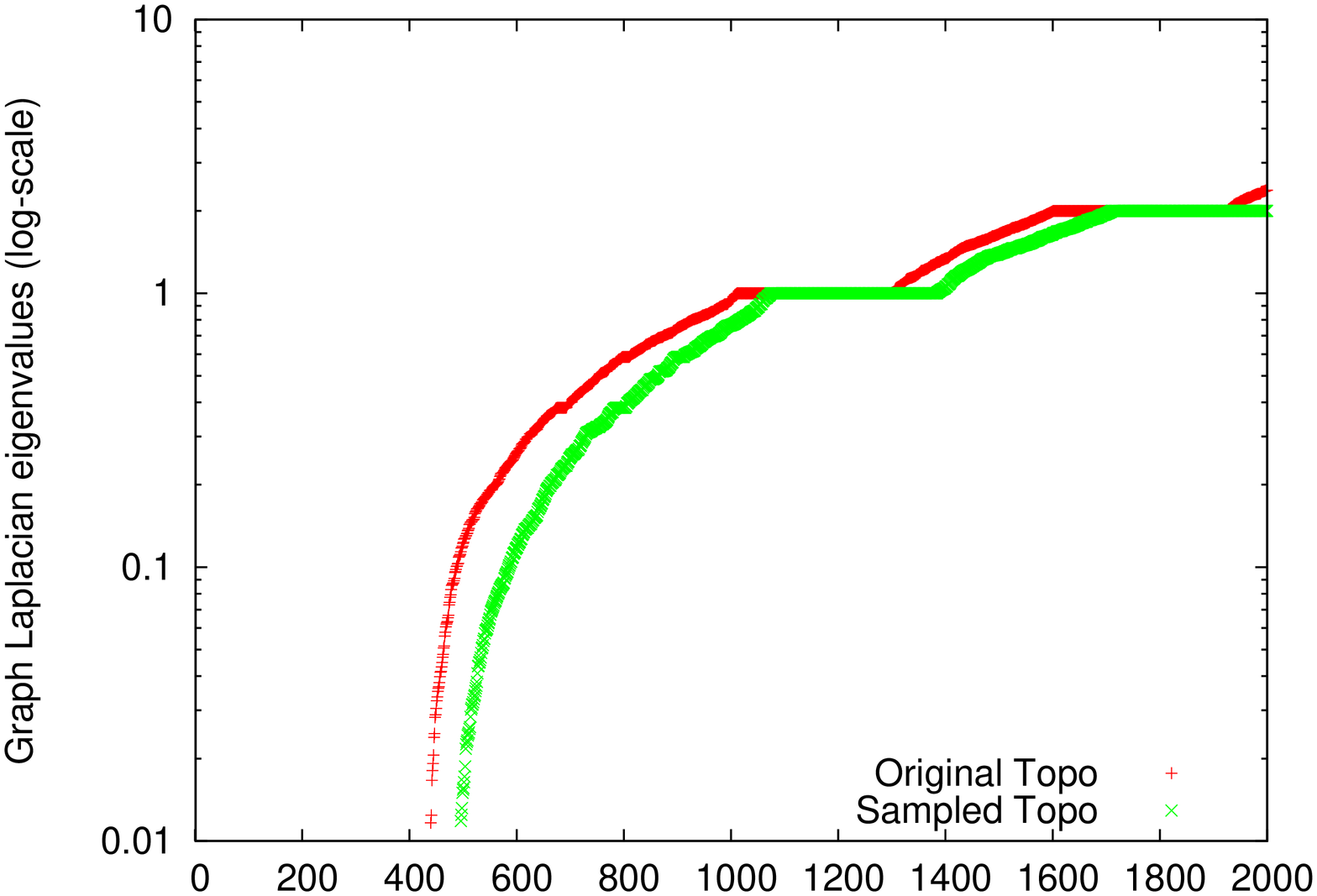}
    }
    \qquad
    \subfloat[Flickr]
    {
        \includegraphics[angle=-90,width=0.45\textwidth]{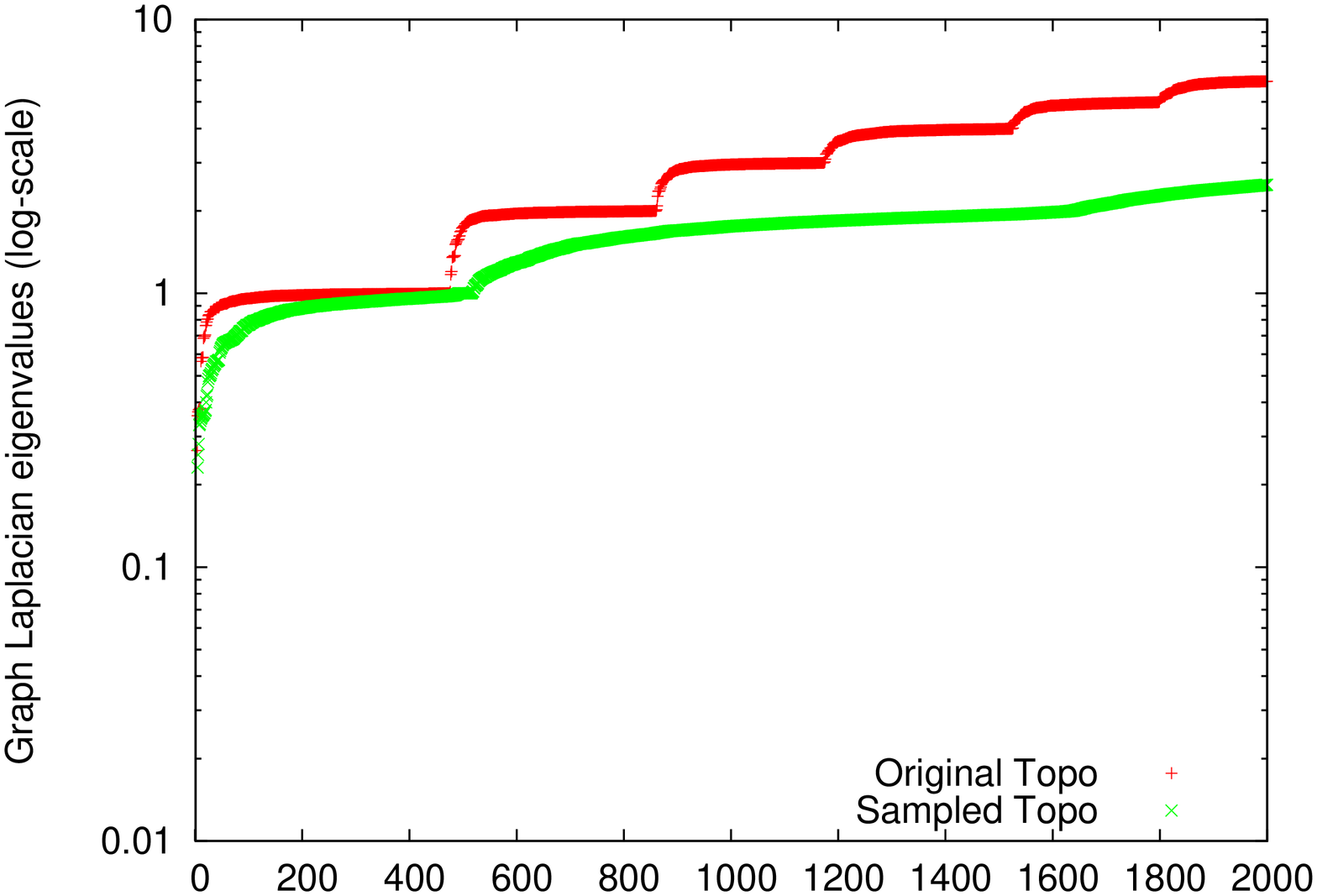}
    }
    \caption{Eigenvalues of graph Laplacian before and after simplification}
    \label{fig:laplacian}
\end{figure}

In this section we investigate the impact of topological simplification (or sampling) on the spectrum of the graph. For both CiteSeer and Flickr (results for Wikipedia are similar to that of Flickr) we compute the Laplacian of the graph and then examine the top part of its eigenspectrum (first 2000 eigenvectors). Specifically, in Figure \ref{fig:laplacian} we order the eigenvectors from the smallest one to the largest one (on the X axis) and plot corresponding eigenvalues (on the Y axis). 

The multiplicity of 0 as an eigenvalue in such a plot corresponds to the number of independent components within the graph \cite{mohar1991laplacian}. For CiteSeer we see an increase in the number of components as a result of topological simplification whereas for Flickr (similarly for Wikipedia) the number of components is unchanged. Our hypothesis is that for datasets like CiteSeer this will have a negative impact on the quality of the resulting clustering. We further hypothesize that our content-based enhancements will help in overcoming this shortfall.  

Note that the sum of eigenvalues for the complete spectrum is proportional to the number of edges in the graph \cite{mohar1991laplacian} so this explains why the plots for the original graphs are slightly above those for the simplified graph even though the overall trends (e.g. spectral gap, relative changes in eigenvalues), except for the number of components, are quite similar for both datasets.

\subsection{Clustering Quality}
\label{subsec:clustering_quality}
%Clustering quality is arguably the most important factor to measure the merit of any clustering algorithm.
We are interested in comparison between the predicted clustering and the real community structure since group/category information is available for all three datasets. Later in Section \ref{sec:casestudy} we will evaluate CODICIL's performance qualitatively. While it is tempting to use conductance or other cut-based objectives to evaluate the quality of clustering, they only value the structural cohesiveness but not the content cohesiveness of resultant clustering, which is exactly the motivation of content-aware clustering algorithm. Instead, we use average F-score with regard to the ground truth as the clustering quality measure, as it takes content grouping into consideration and ensures a fair comparison among different clusterings.
%To ensure different clusterings' qualities are comparable, average F-score is used as the measure of clustering quality, and it is computed in the following way.
Given a predicted cluster $p$ and with reference to a ground truth cluster $g$ (both in the form of node set), we define the precision rate as $\frac{|p \cap g|}{|p|}$ and the recall rate as $\frac{|p \cap g|}{|g|}$. The F-score of $p$ on $g$, denoted as $F(p,g)$, is the harmonic mean of precision and recall rates. 

For a predicted cluster $p$, we compute its F-score on each $g$ in the ground truth clustering $G$ and define the maximal obtained as $p$'s F-score on $G$. That is:
\begin{equation}
    F(p,G) = \max_{g \in G}{F(p,g)} \enspace .
\end{equation}

The final F-score of the predicted clustering $P$ on the ground truth clustering $G$ is then calculated as the weighted (by cluster size) average of each predicted cluster's F-score:
\begin{equation}
    F(P,G) = \sum_{p \in P}{\frac{|p|}{|\mathcal{V}|} F(p,G)} \enspace .
\end{equation}
This effectively penalizes the predicted clustering that is not well-aligned with the ground truth, and we use it as the quality measure of all methods on all datasets.

\subsubsection{CiteSeer}
\begin{figure}[htb]
    \begin{center}
        \includegraphics[angle=-90,width=0.45\textwidth]{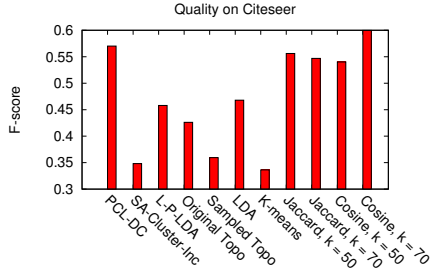}
    \end{center}
    \caption{F-score of Metis on CiteSeer}
    \label{fig:fscore_citeseer}
\end{figure}

In Figure~\ref{fig:fscore_citeseer} we show the experiment results on CiteSeer. Since it is known that the network has six communities (i.e. sub-fields in computer science), there is no need to vary $l$, the number of desired clusters. We report results using Metis (similar numbers were observed with Markov clustering)\footnote. For PCL-DC, we set the parameter $\lambda$ to 5 as suggested in the original paper, yielding an F-score of 0.570. The F-scores of SA-Cluster-Inc and L-P-LDA are 0.348 and 0.458, respectively. As we can see clearly in the bar chart, clustering based on topology alone results in a performance well below the state-of-the-art content-aware clustering methods. This is not surprising as the input graph has 438 connected components and therefore most small components were randomly assigned a prediction label. Although such approach has no impact on topology-based measures (e.g. normalized cut or conductance), it greatly spoils the F-score measure against the ground truth. Neither is LDA able to provide a competitive result, as it is oblivious to link structure embedded in the dataset. Surprisingly though, K-means only manages to produce a very unbalanced clustering (the largest cluster always contains more than 90\% of all papers) even after 50 iterations, and its F-score (averaged over five runs) is only 0.336.

On the other hand, our content-aware approaches (using Metis as the clustering method) were able to handle the issue of disconnection as they also include content-similar edges. For both similarity measures, the F-scores are within 90\% range of PCL-DC, and it outperforms PCL-DC when $k$ increases to 70.
%When $k=50$, the F-score of zNorm is within the 90\% range of PCL-DC and it becomes higher when $k$ is increased to 70.
%of For full-graph content similarity search, ``zNorm'' is below PCL-DC by only 0.005 while ``Max'' has an improvement of 3\% over PCL-DC. The improvement provided by full graph search over both 1-hop and 2-hop search is more than 30\%. We attribute this gain to the fact that CiteSeer's graph density is very low, and only with full-graph search are we able to discover a sufficient number of content-similar documents. 

While achieving the quality that is comparable with existing methods, the CODICIL series are significantly faster. PCL-DC takes 234 seconds on this dataset and SA-Cluster-Inc requires 306 seconds. LDA finishes in 40 seconds. In contrast, the sum of CODICIL's edge sampling and clustering time never exceeds 1 second.
%The one-time cost of content graph construction is 6 seconds.
Therefore, the CODICIL methods are at least one order of magnitude faster than state-of-the-art algorithms. 
\subsubsection{Wikipedia}
For the Wikipedia dataset, we were unable to run the experiment on SA-Cluster-Inc, PCL-DC, L-P-LDA, LDA and K-means as their memory and/or running time requirement became prohibitive on this million-node network. For example, storing 10,000 centroids alone in K-means requires 54 GBs).

\begin{figure*}[htb]
    \centering
    \subfloat[F-score of MLR-MCL on Wikipedia]
    {
    \includegraphics[angle=-90,width=0.45\textwidth]{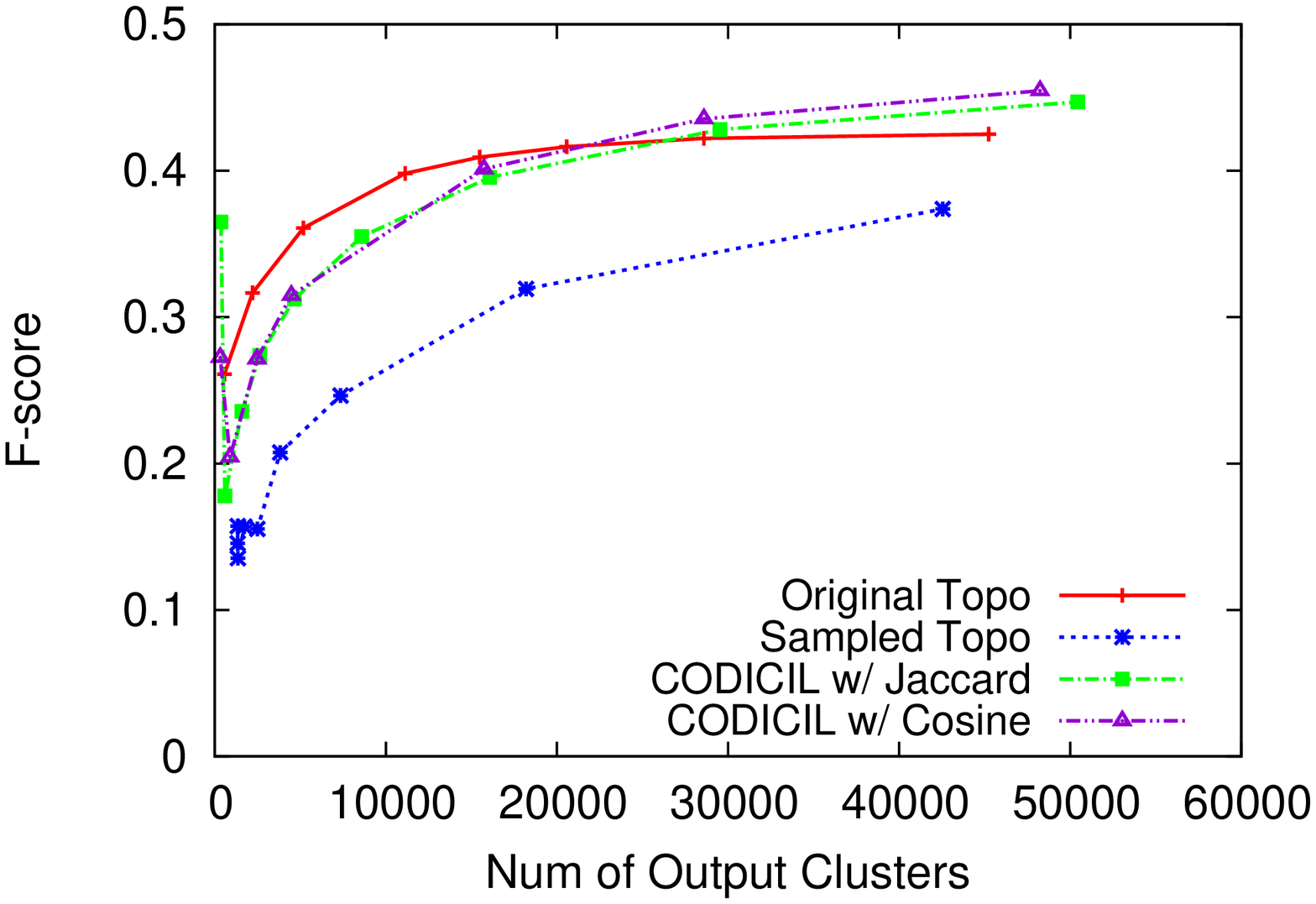}
    \label{fig:fscore_wikipedia_mlrmcl}
    }
    \qquad
    \subfloat[Running time of MLR-MCL on Wikipedia]
    {
    \includegraphics[angle=-90,width=0.45\textwidth]{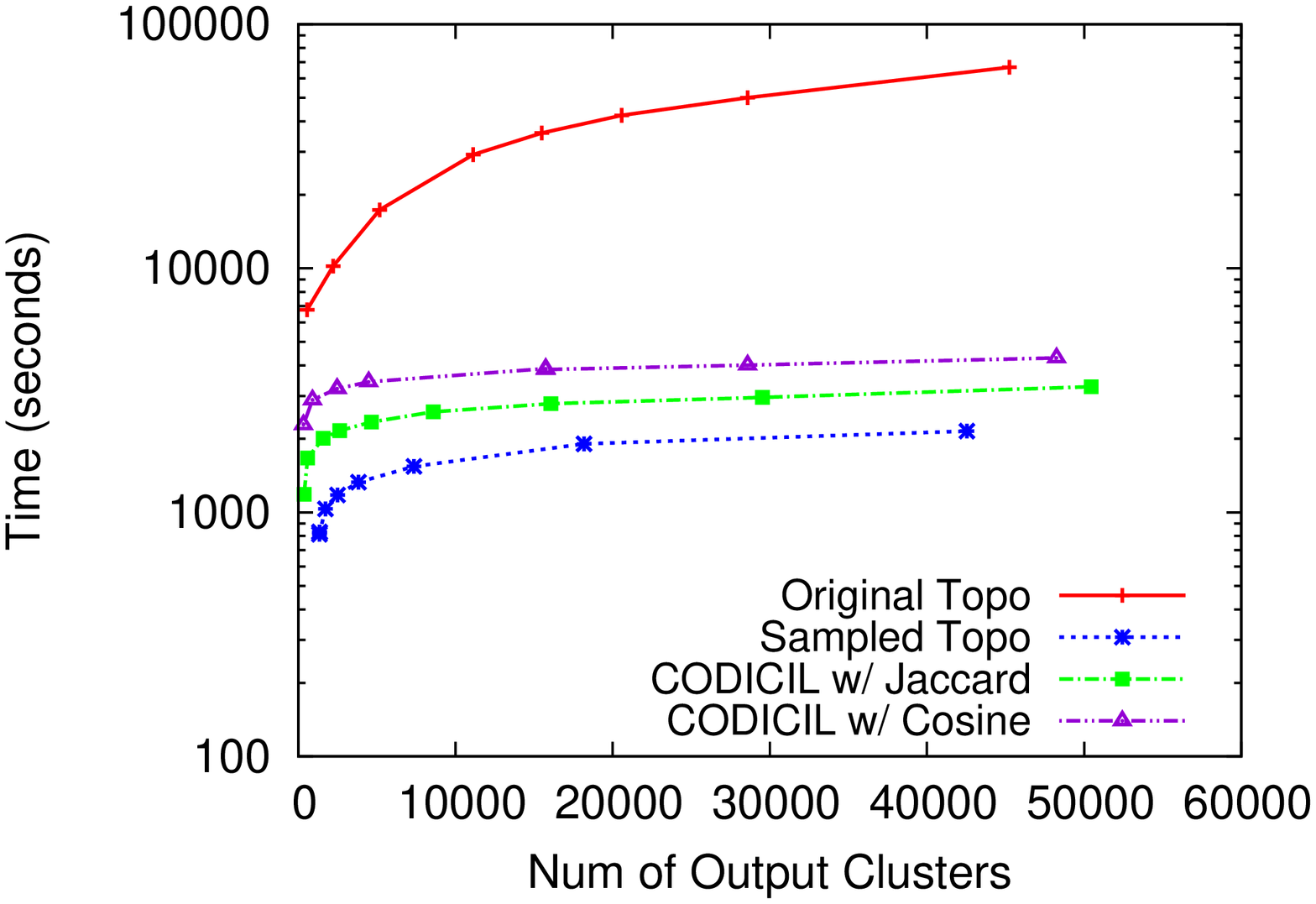}
    \label{fig:time_wikipedia_mlrmcl}
    }
    \qquad
    \subfloat[F-score of Metis on Wikipedia]
    {
    \includegraphics[angle=-90,width=0.45\textwidth]{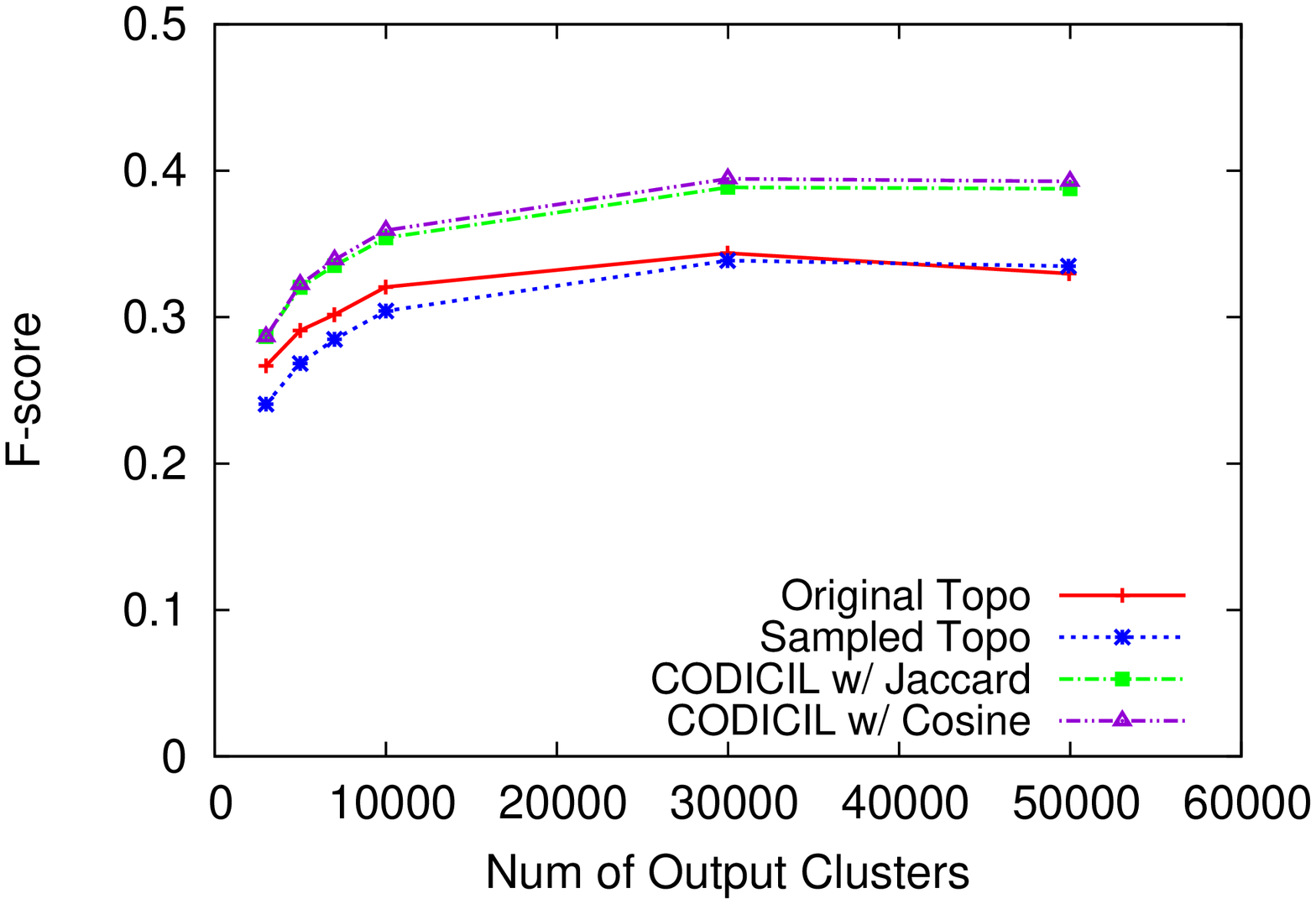}
    \label{fig:fscore_wikipedia_metis}
    }
    \qquad
    \subfloat[Running time of Metis on Wikipedia]
    {
    \includegraphics[angle=-90,width=0.45\textwidth]{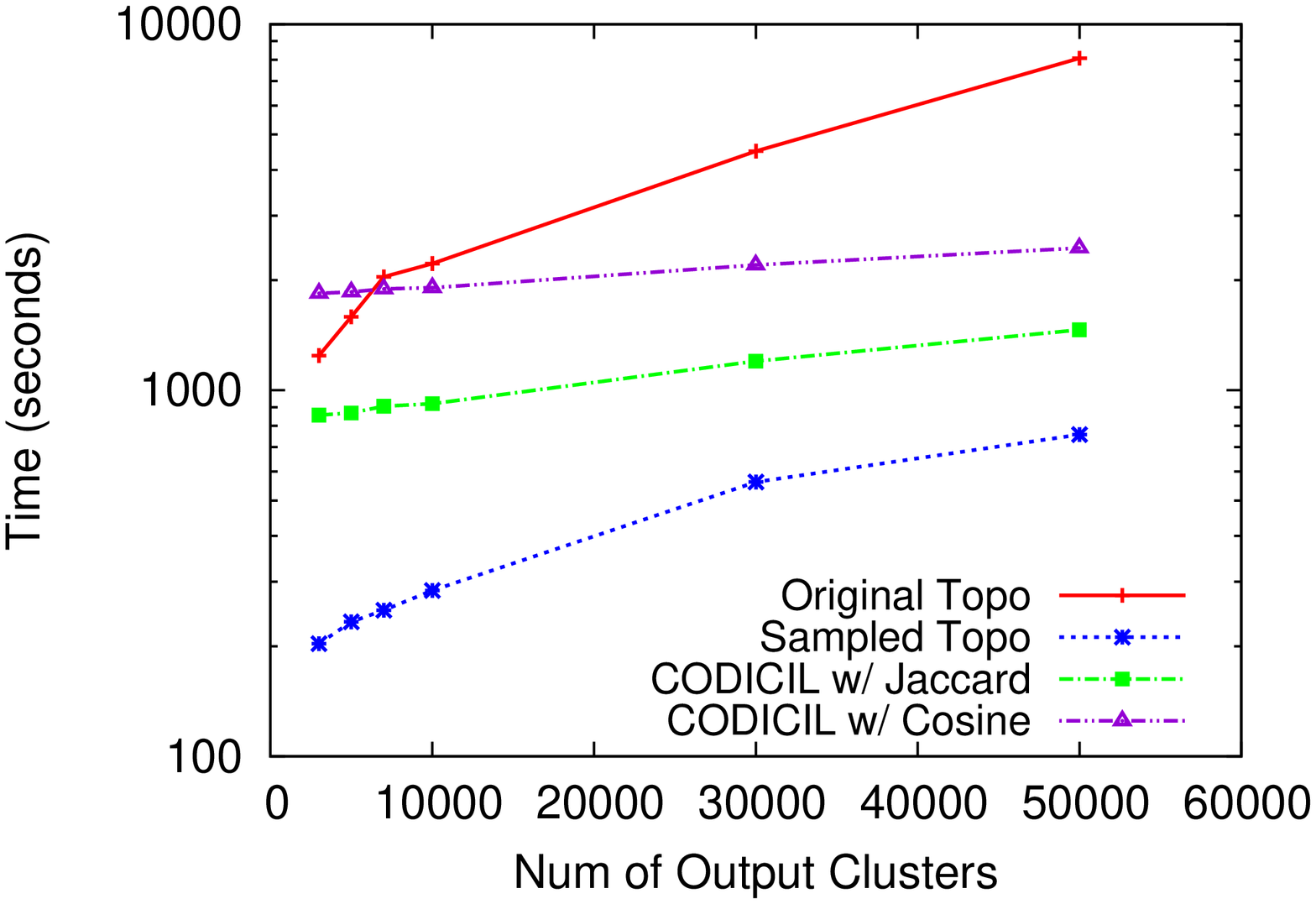}
    \label{fig:time_wikipedia_metis}
    }
    \label{fig:wikipedia_results}
    \caption{Experiment Results on Wikipedia}
\end{figure*}

Figures \ref{fig:fscore_wikipedia_mlrmcl} and \ref{fig:fscore_wikipedia_metis} plot the performances using MLR-MCL and Metis, respectively. Since category assignments as the ground truth are overlapping, there is no gold standard for the number of clusters. We therefore varied $l$ in both clustering algorithms. Our content-aware clustering algorithms constantly outperforms Sampled Topo by a large margin, indicating that CODICIL methods are able to simplify the network and recover community structure at the same time. CODICIL methods' F-scores are also on par or better than those of Original Topo.
%For Metis, the quality of our content-aware clustering algorithms always exceed that of clustering using network structure only. For MLR-MCL, the benefit of adding content information also increases steadily when the clustering is sufficiently refined.
%The comparison among 1-hop, 2-hop and full graph content similarity search is also demonstrated on the plots. The advantages of 2-hop and full graph search shows our algorithm's capability of grouping documents whose relationship would otherwise be overlooked.
%For all categories which contain at least 10 documents, the average document number is 84, making 40,000 to 50,000 a reasonable range for number of clusters. Furthermore, zNorm is always close to the max score, signaling one advantage of zNorm method, as it avoid the potentially expensive step-wise search for the optimal parameter. 
%In Figure~\ref{fig:fscore_wikipedia_metis_cGraph}, we compare the results of clustering algorithms for which different content similarity search radii are applied. 
\subsubsection{Flickr}
\begin{figure*}[htb]
    \centering
    \subfloat[F-score of MLR-MCL on Flickr]
    {
    \includegraphics[angle=-90,width=0.45\textwidth]{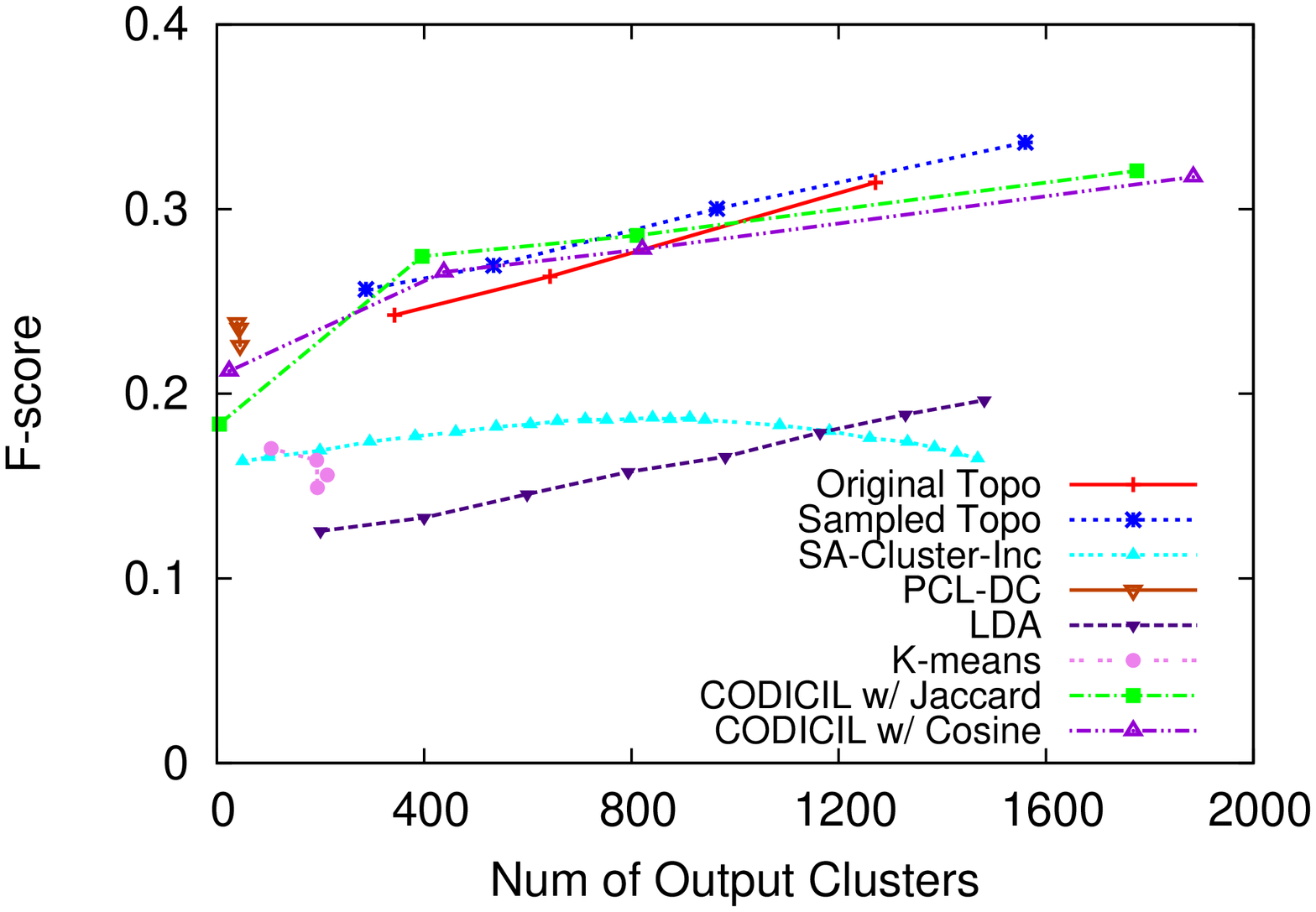}
    \label{fig:fscore_flickr_mlrmcl}
    }
    \qquad
    \subfloat[Running time of MLR-MCL on Flickr]
    {
    \includegraphics[angle=-90,width=0.45\textwidth]{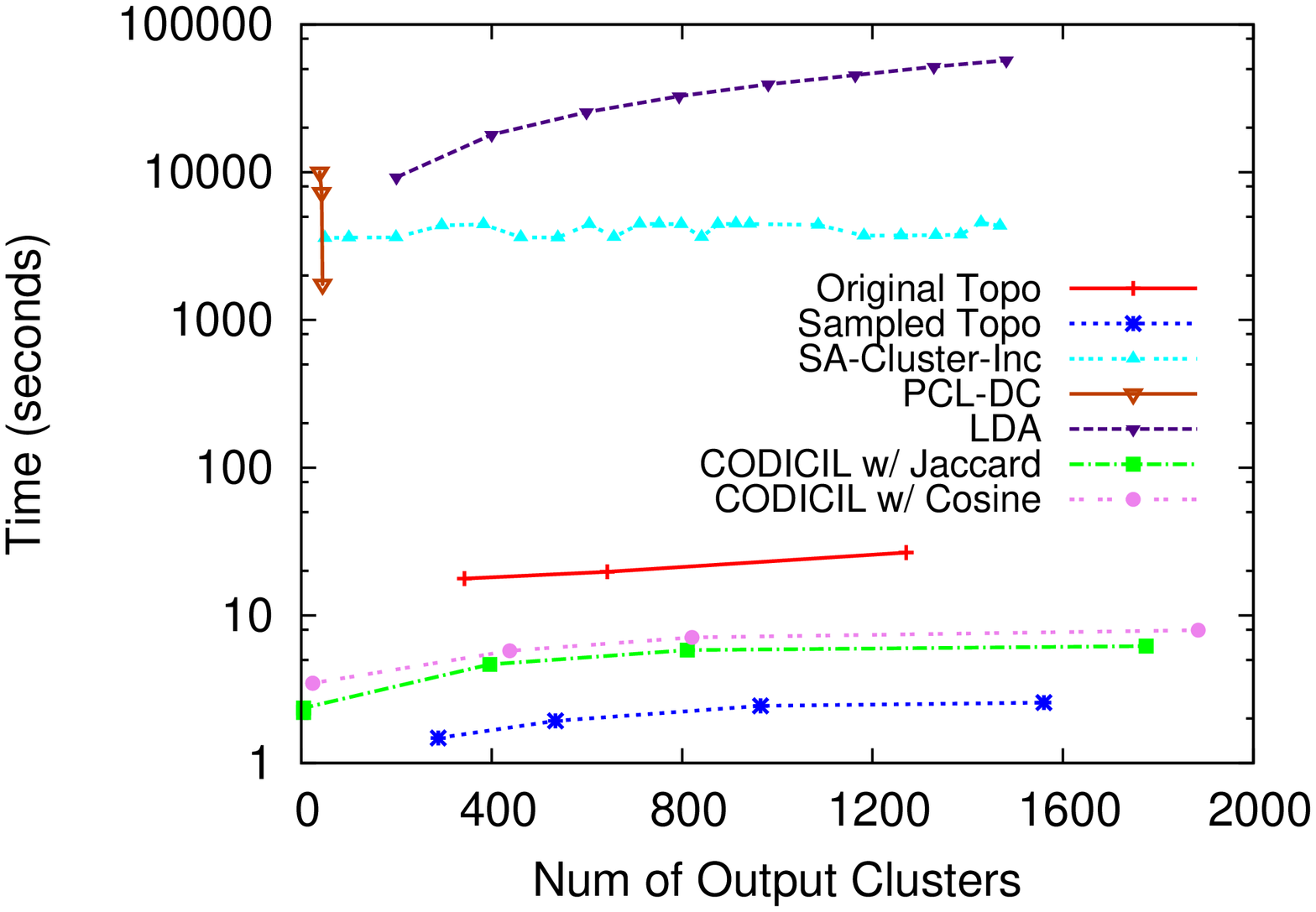}
    \label{fig:time_flickr_mlrmcl}
    }
    %\qquad
    %\subfloat[F-score of Metis on Flickr]
    %{
    %\includegraphics[angle=-90,width=0.45\textwidth]{plot/Flickr/metis}
    %\label{fig:fscore_flickr_metis}
    %}
    %\qquad
    %\subfloat[Running time of Metis on Flickr]
    %{
    %\includegraphics[angle=-90,width=0.45\textwidth]{plot/Flickr/metis_time}
    %\label{fig:time_flickr_metis}
    %}
    \label{fig:flickr_results}
    \caption{Experiment Results on Flickr}
\end{figure*}

%Figures \ref{fig:fscore_flickr_mlrmcl} and \ref{fig:fscore_flickr_metis} show the performances of various methods on Flickr,
Figure \ref{fig:fscore_flickr_mlrmcl} shows the performances of various methods with MLR-MCL on Flickr, 
where SA-Cluster-Inc, PCL-DC, LDA and K-means can also finish in a reasonable time (L-P-LDA still takes more than 30 hours).
Again, $l$ was varied for the clustering algorithm. Similar to results on CiteSeer, CODICIL methods again lead the baselines by a considerable margin. The F-scores of SA-Cluster-Inc, LDA, and K-means never exceed 0.2, whereas CODICIL methods' F-scores are often higher, together with Original \& Sampled Topo.
%Again, $l$ was varied for both clustering algorithms. Similar to results on CiteSeer, CODICIL methods again lead the baselines by a considerable margin. The F-scores of both SA-Cluster-Inc and LDA never exceed 0.2, whereas CODICIL methods' F-scores are always above 0.3. Topology-only methods (Original \& Sampled Topo) also underperform CODICIL in general.

Readers may have noticed that for PCL-DC only three data points ($l=50,75,100$) are obtained. That is because its excessive memory consumption crashed our workstation after using up 16 GBs of RAM for larger $l$ values. We also observe that while PCL-DC generates a group membership distribution over $l$ groups for each vertex, fewer than $l$ communities are discovered. That is, there exist groups of which no vertex is a prominent member. Furthermore, the number of communities discovered is decreasing as $l$ increases (45, 43 and 39 communities for $l=50,75,100$), which is opposite to other methods' trends. All three clusterings' F-scores are less than 0.25. Similarly, multiple runs of K-means (K is set to 400, 800, 1200, and 1600) can only identity roughly 200 communities.
\subsection{Scalability}
\label{scalability}
%We discuss the algorithms' running time and spatial cost in this subsection. A strong dependency of temporal/spatial cost on graph size is a critical issue as the graph reaches medium to large scale. Furthermore, if the pre-specified number of clusters has an impact on algorithm cost, the overhead will also becomes considerable since often does the cluster count naturally increase as the network of interest grows. 
The running time on CiteSeer has already been discussed, and here we focus on Flickr and Wikipedia. For CODICIL methods, the running time includes both edge sampling and clustering stage. The plots' Y-axes (running time) are in log scale.
\subsubsection{Flickr}
We first report scalability results on Flickr
%(see Figures~\ref{fig:time_flickr_mlrmcl} and~\ref{fig:time_flickr_metis}).
(see Figure~\ref{fig:time_flickr_mlrmcl}).
For SA-Cluster-Inc, the value of $l$ (the desired output cluster count), ranging from 100 to 5000, does not affect its running time as it always stays between 1 and 1.25 hours with memory usage around 12GB. The running time of LDA appears, to a large extent, linear in the number of latent topics (i.e. $l$) specified, climbing up from 2.56 hours ($l=200$) to 15.88 hours ($l=1600$). For PCL-DC, the running time with three $l$ values ($50,75,100$) is 0.5, 2.0 and 2.8 hours, respectively.
%Note that for all clustering algorithms the number of clusters generated does not necessarily equal the specified value.
%While it takes the program 0.5 hours to generate 50 clusters, the running time drastically increases to 2.8 hours as the cluster count is changed to 100.

As for our content-aware clustering algorithms, running them on Flickr requires less than 8 seconds, which is three to four orders of magnitude faster than SA-Cluster-Inc, PCL-DC and LDA. Original Topo takes more than 10 seconds, and Sampled Topo runs slightly faster than CODICIL methods. %at the cost of a lower quality.
%\begin{table}
    %\hspace{-1.7cm}
    %\small
    %\centering
    %\begin{tabular}{|c|c|c|c|}
        %\hline
        %& $\mathcal{G}_c$ Construction & Sampling & Clustering (Metis) \\ \hline
        %Original Topo. & - & - & 4.76 \\ \hline
        %Sampled Topo. & - & 0.24 & 1.51 \\ \hline
        %Full Search, zNorm & 135 & 0.56 & 1.21  \\ \hline
        %2-Hop, zNorm & 980 & 0.57 & 1.20 \\ \hline
        %1-Hop, zNorm & 11 & 0.46 & 1.05  \\ \hline
    %\end{tabular}
    %\caption{Running time (in seconds) of CODICIL and topology-only methods on Flickr. Metis with 1000 clusters}
    %\label{tab:time_flickr}
%\end{table}
%Table \ref{tab:time_flickr} shows the running time breakdown for the proposed CODICIL algorithms as well as conventional content-insensitive approaches. Again we note that building content graph is only a one-time cost as all edge sampling processes will use the same content graph. Furthermore, the construction process can be easily parallized since it operates on each graph node separately.\\
%The running time of CODICIL algorithms versus conventional content-insensitive methods for Flickr is plotted in Figure~\ref{fig:time_flickr_mlrmcl} and~\ref{fig:time_flickr_metis}. Very similar trend is seen from them, as CODICIL methods are many times faster 

%For the Wikipedia dataset, we run all experiments on a workstation equipped with dual-socket, quad-core 2.5 GHz Opteron processors and 24 GB of physical memory. Neither SA-Cluster-Inc nor PCL-DC can handle such large dataset within the memory budget, while other methods including all CODICIL ones finished successfully.
\subsubsection{Wikipedia}
Original Topo, Sampled Topo and all CODICIL methods finished successfully. The running time is plotted in Figures~\ref{fig:time_wikipedia_mlrmcl} and~\ref{fig:time_wikipedia_metis}. When clustering using MLR-MCL, our methods are at least one order of magnitude faster than clustering based on network topology alone. For Metis, CODICIL is also more than four times faster. The trend lines suggest our methods have promising scalability for analysis on even larger networks.
\subsection{Effect of Varying $\alpha$ on F-score }
\label{sec:vary_alpha}
So far all experiments performed fix $\alpha$ at 0.5, meaning equal weights of structural and content similarities. In this sub-section we track how the clustering quality changes when the value of $\alpha$ is varied from 0.1 to 0.9 with a step length of 0.1.
\begin{figure*}[htb]
    \centering
    \subfloat[Varying $\alpha$ on Wikipedia]
    {
        \includegraphics[angle=-90,width=0.33\textwidth]{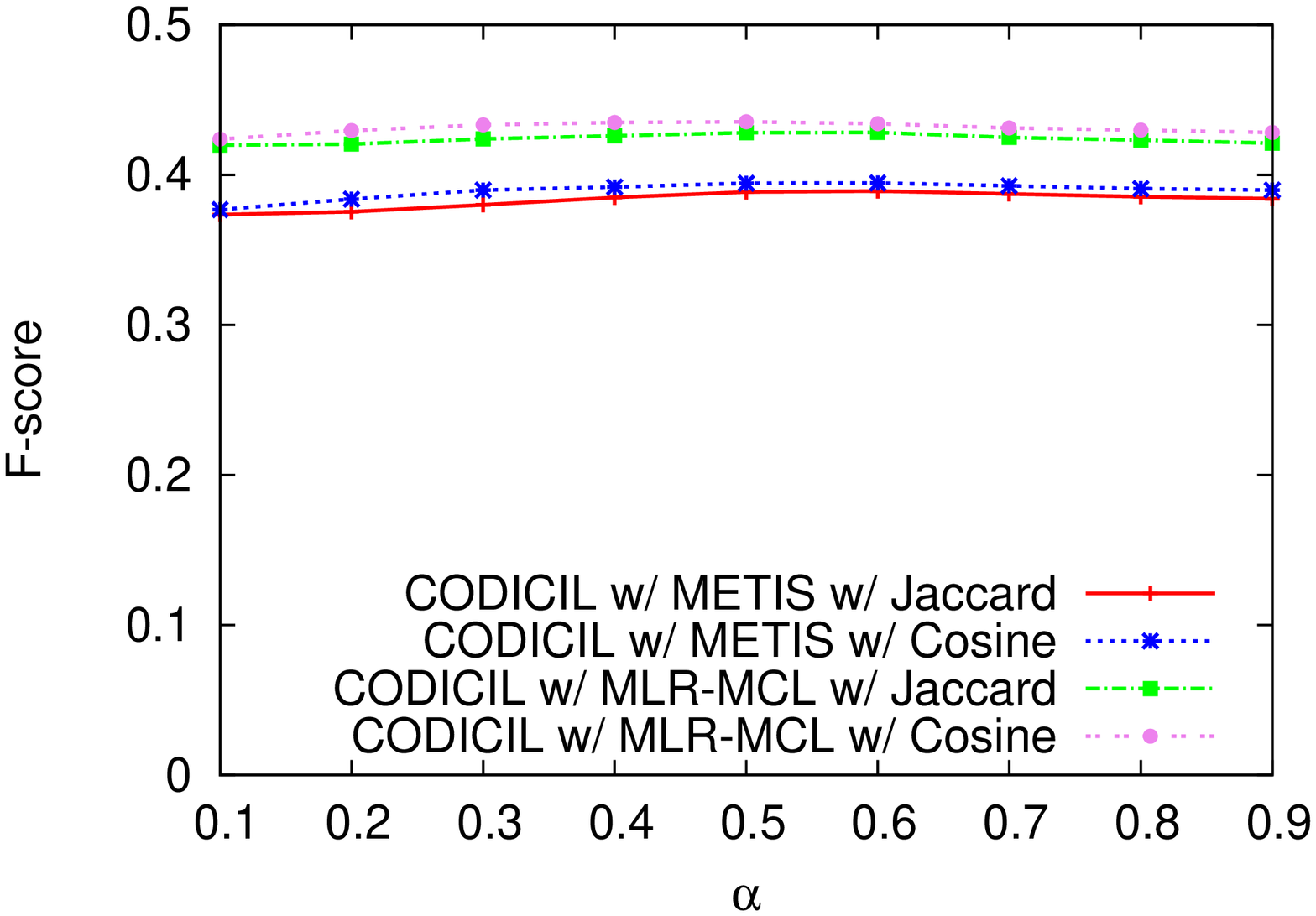}
        \label{fig:wikipedia_alpha}
    }
    %\qquad
    \subfloat[Varying $\alpha$ on Citeseer]
    {
        \includegraphics[angle=-90,width=0.33\textwidth]{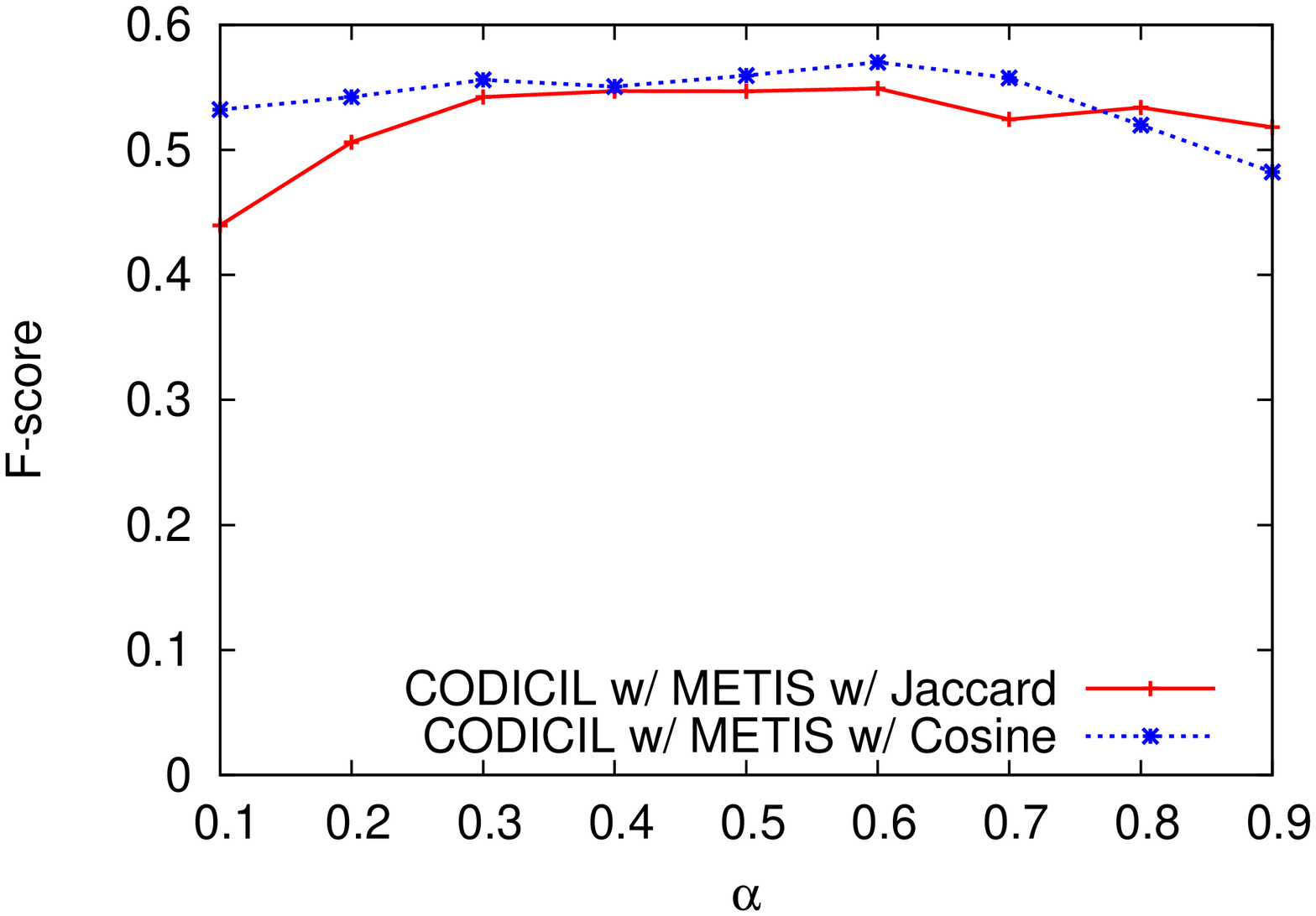}
        \label{fig:citeseer_alpha}
    }
    %\qquad
    \subfloat[Varying $\alpha$ on Flickr]
    {
        \includegraphics[angle=-90,width=0.33\textwidth]{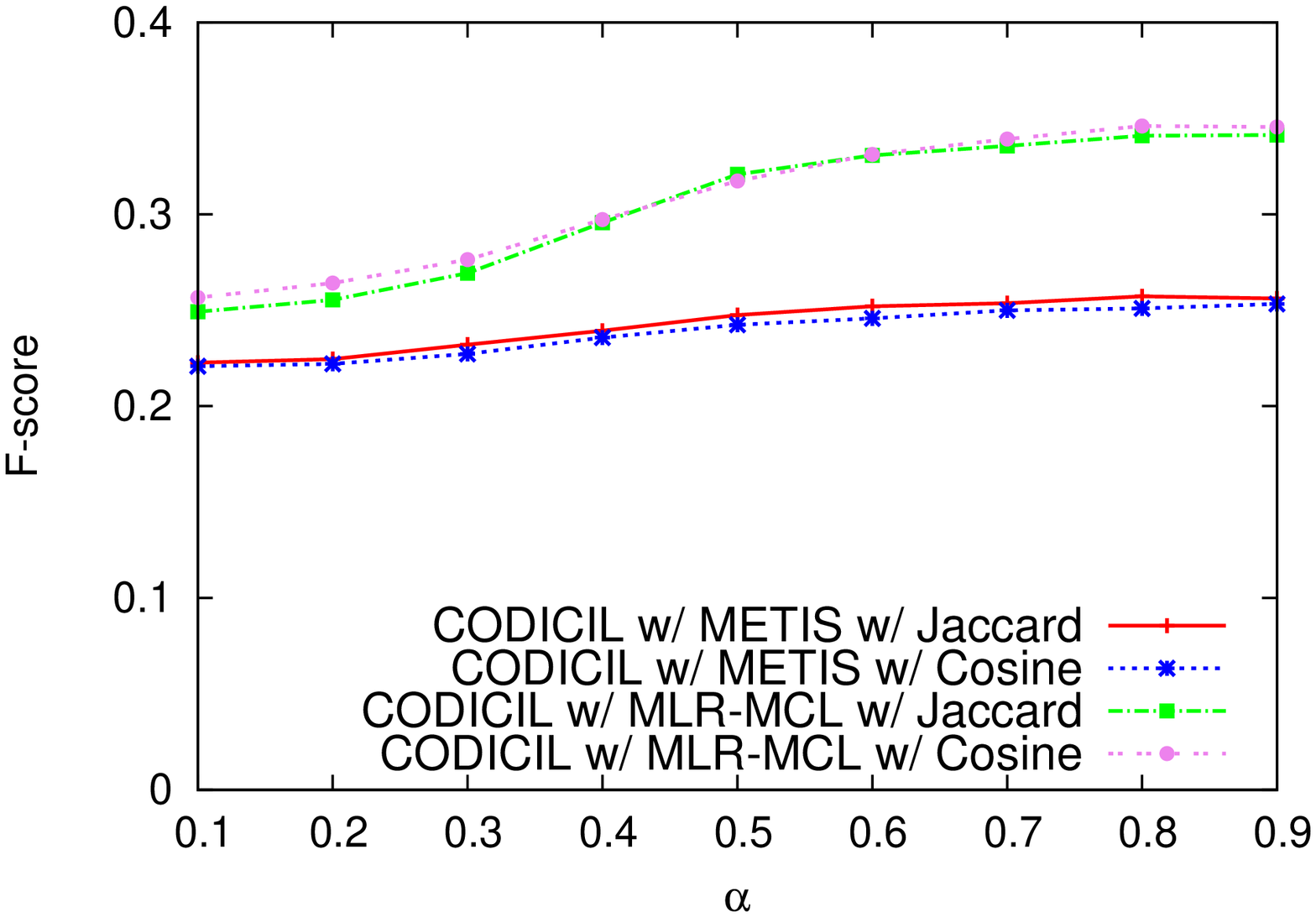}
        \label{fig:flickr_alpha}
    }
    \label{fig:alpha}
    \caption{Effect of Varying $\alpha$ on F-score (Avg. \# Clusters for Wikipedia: 29,414, Avg. \# Clusters for Flickr: 1,911)} 
\end{figure*}

On Wikipedia (Figure \ref{fig:wikipedia_alpha}) and Citeseer (Figure \ref{fig:citeseer_alpha}), F-scores are greatest around $\alpha=0.5$, supporting the decision of assigning equal weights to structural and content similarities. Results differ on Flickr where F-score is constantly improving when $\alpha$ increases (i.e. more weight assigned to topological similarity).
%This shows the flexibility of CODICIL in handling networks with strong topology signals.
%This explains the absence of clear advantage of CODICIL over topology-only clustering methods on Flickr dataset.
\subsection{Effect of \ensuremath{\mathcal{E}_c} Constraint on F-score}
In Section \ref{sec:m} we discuss the possibility of constraining content edges within a topological neighborhood for each node $v_i$. Here we provide a brief review on how the qualities of resultant clusterings are impacted by such constraint. For the sake of space, we focus on the F-scores on Wikipedia and Flickr.

\begin{figure*}[htb]
    \centering
    \subfloat[Varying $\mathcal{E}_c$ Constraint, MLR-MCL on Wiki.]
    {
    \includegraphics[angle=-90,width=0.45\textwidth]{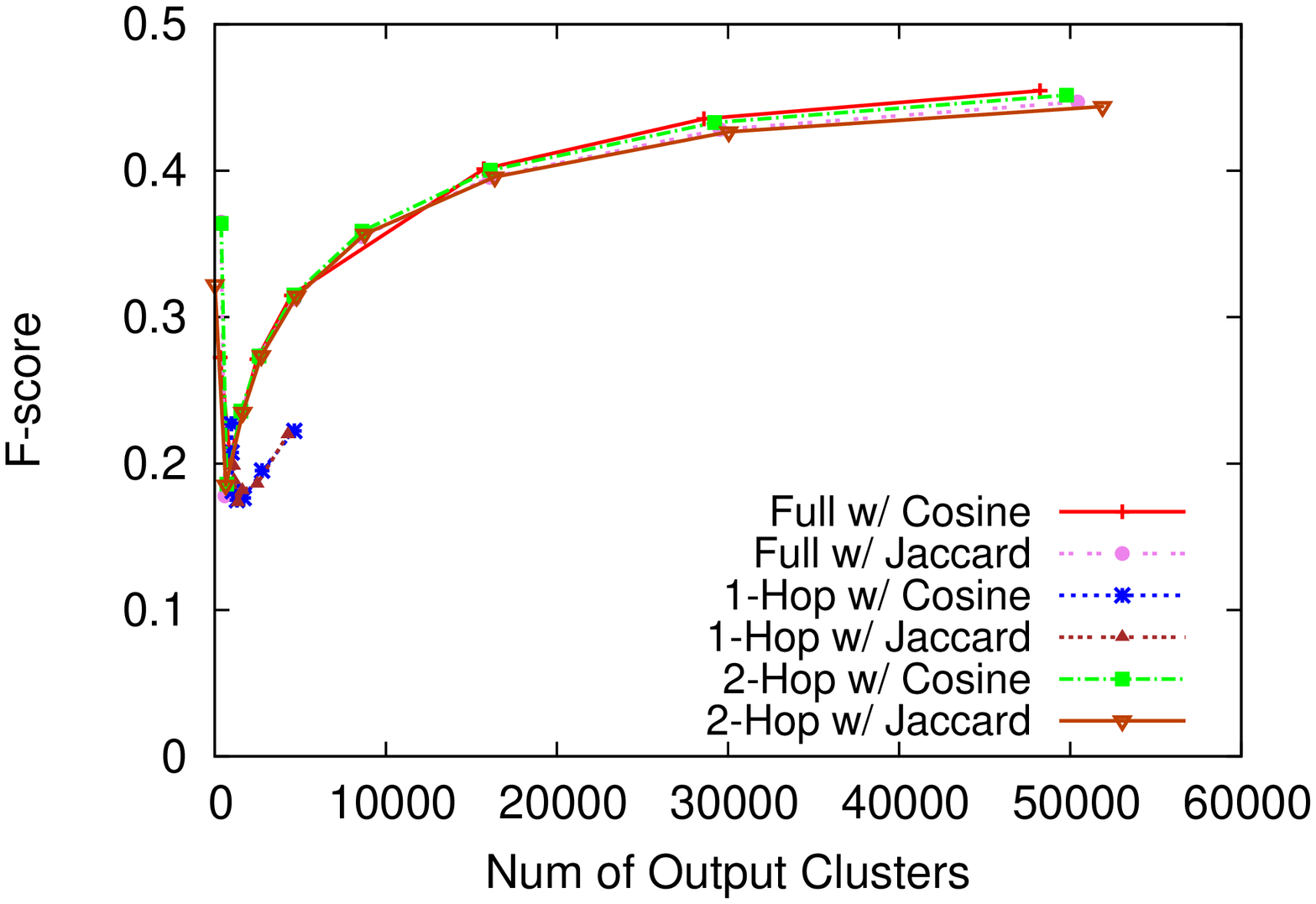}
    \label{fig:wikipedia_mlrmcl_cGraph}
    }
    \qquad
    \subfloat[Varying $\mathcal{E}_c$ Constraint, METIS on Wiki.]
    {
    \includegraphics[angle=-90,width=0.45\textwidth]{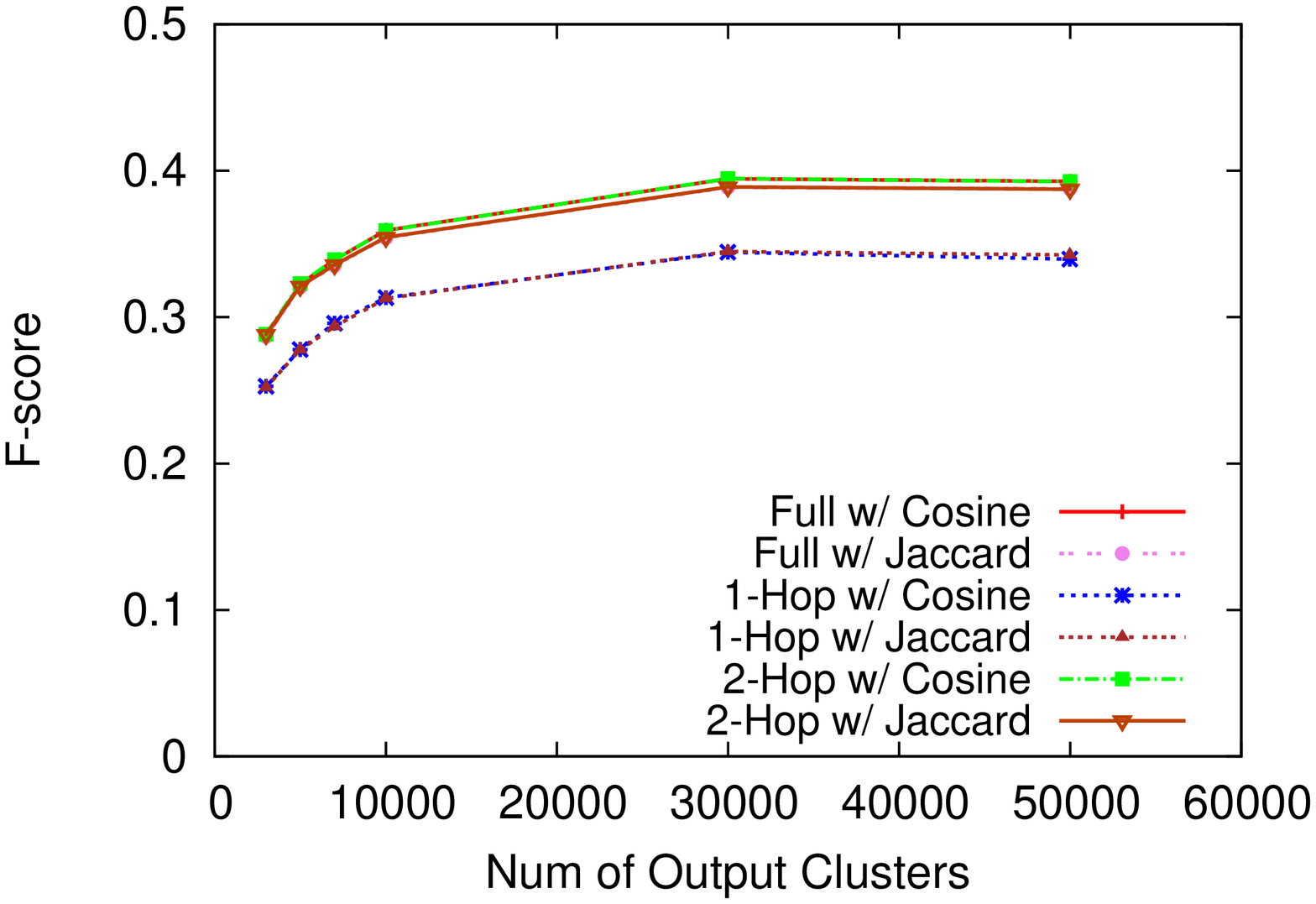}
    \label{fig:wikipedia_metis_cGraph}
    }
    \qquad
    \subfloat[Varying $\mathcal{E}_c$ Constraint, MLR-MCL on Flickr]
    {
    \includegraphics[angle=-90,width=0.45\textwidth]{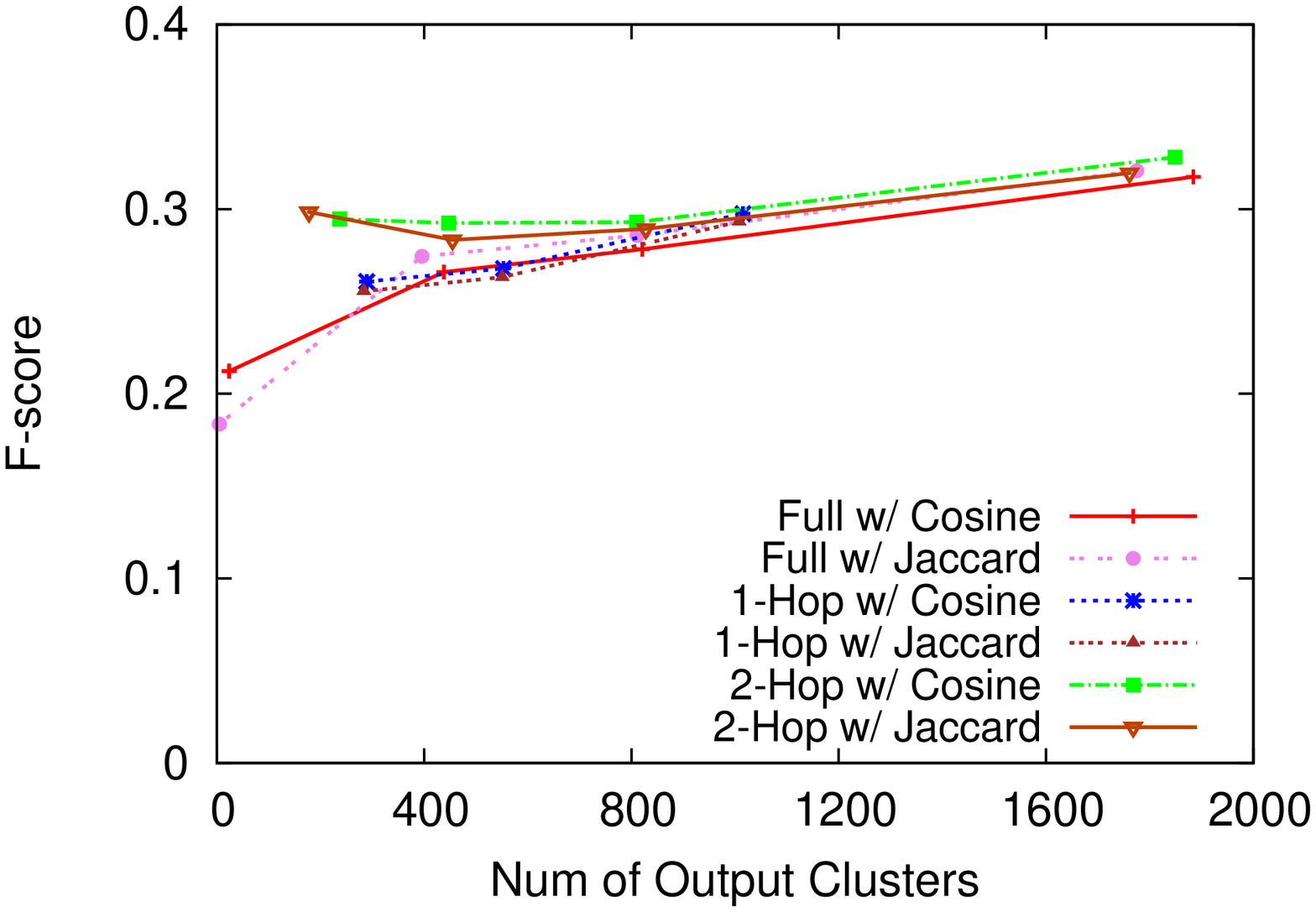}
    \label{fig:flickr_mlrmcl_cGraph}
    }
    \qquad
    \subfloat[Varying $\mathcal{E}_c$ Constraint, METIS on Flickr]
    {
    \includegraphics[angle=-90,width=0.45\textwidth]{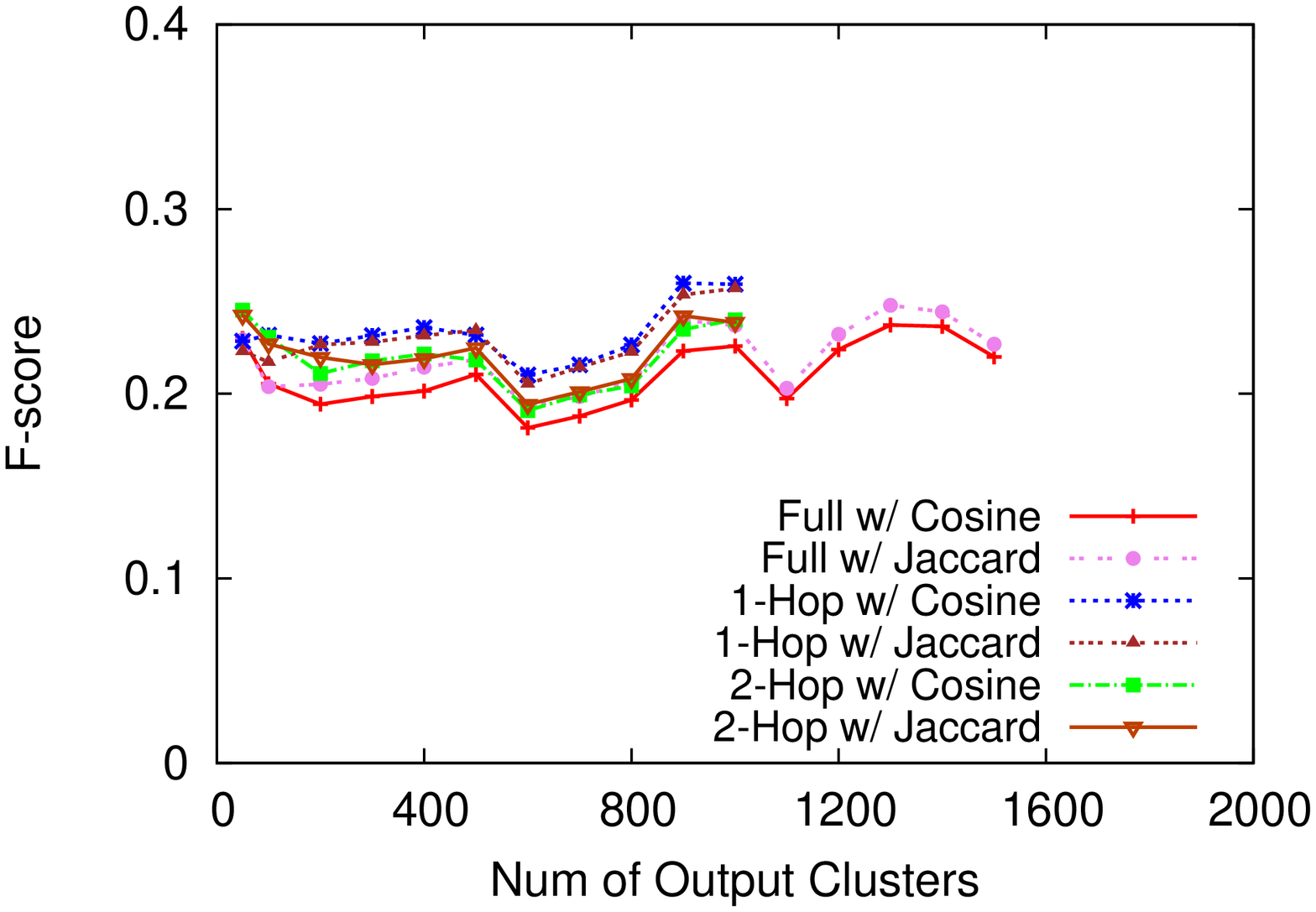}
    \label{fig:flickr_metis_cGraph}
    }
    \label{fig:cGraph}
    \caption{Effect of \ensuremath{\mathcal{E}_c} Constraint on F-score}
\end{figure*}

Figures \ref{fig:wikipedia_mlrmcl_cGraph} and \ref{fig:wikipedia_metis_cGraph} show F-scores achieved on Wikipedia, using different $\mathcal{E}_c$ constraints. \emph{Full} means no constraint and thus $TopK$ sub-routine searches the whole vertex set $\mathcal{V}$, whereas \emph{1-hop} constrains the search to within a one-hop neighborhood, and likewise for \emph{2-hop}.
%for content neighbors have been defined in Section \ref{sec:m}.
The plots of \emph{full} and \emph{2-hop} almost overlap with each other, suggesting that searching within the 2-hop neighborhood can provide sufficiently strong content signals on this dataset.
For Flickr (Figures \ref{fig:flickr_mlrmcl_cGraph} and \ref{fig:flickr_metis_cGraph}), interestingly \emph{2-hop} and \emph{1-hop} have a slight lead over \emph{full}. This may be an indication that in online social networks, compared with information networks, content similarity between two closely connected users emits stronger community signals.
%While its indication requires further investigation, we speculate that in online social networks content similarity between two close-knit users emits stronger community signals.
%For Flickr (Figures \ref{fig:flickr_mlrmcl_cGraph} and \ref{fig:flickr_metis_cGraph}) \emph{full} still leads \emph{2-hop} to some extent. On both datasets, the gap between performances of \emph{1-hop} and \emph{full} is considerable, indicating the trade-off between running time and amount of content signals recovered.
\subsection{Discussions}
An interesting observation on the biased edge sampling is that it always results in an improvement in running time. However, sampling just the topology graph results in a clear loss in accuracy whereas content-conscious sampling is much more effective with accuracies that are on par with the best performing methods at a fraction of the cost to compute. We observe this for all three datasets.

We also find that for probabilistic-model-based methods (PCL-DC, L-P-LDA and LDA) as well as K-means, their running time is at least linear in $l$, the desired number of output clusters, which becomes a critical drawback in face of large-scale workloads. As the network grows, the number of clusters also increases naturally. Plots on CODICIL methods' running time, on the other hand, suggest a logarithmic increase with regard to the number of clusters, which is more affordable.

\section{Case Studies}
\label{sec:casestudy}
In this section, we demonstrate the benefits of leveraging content information on two Wikipedia pages: ``Machine Learning'' and ``Graph (Mathematics)''.

In the original network, ``machine learning'' has a total degree of 637, and many neighbors (including ``1-2-AX working memory task'', ``Wayne State University Computer Science Department'', ``Chou-Fasman method'', etc.) are at best peripheral to the context of machine learning. When we sample the graph according to its link structure only, 119 neighbors are retained for ``machine learning''. Although this eliminates some noise, many others, including the three entries above, are still preserved. Moreover, it also removes during the process many neighbors which should have been kept, e.g. ``naive Bayes classifier'', ``support vector machine'', and so on.

The CODICIL framework, in contrast, alleviates both problems. Apart from removing noisy edges, it also keeps the most relevant ones. For example, ``AdaBoost'', ``ensemble learning'', ``pattern recognition'' all appear in ``machine learning'''s neighborhood in the sampled edge set $\mathcal{E}_{sample}$. Perhaps more interestingly, we find that CODICIL adds ``neural network'', an edge absent from the original network, into $\mathcal{E}_{sample}$ (recall that it is possible for CODICIL to include an edge even it is not in the original graph, given its content similarity is sufficiently high). This again illustrates the core philosophy of CODICIL: to complement the original network with content information so as to better recover the community structure.

Similar observations can be made on the ``Graph (Mathematics)'' page. For example, CODICIL removes entries including ``Eric W. Weisstein'', ``gadget (computer science)'' and ``interval chromatic number of an ordered graph''. It also keeps ``clique (graph theory)'', ``Hamiltonian path'', ``connectivity (graph theory)'' and others, which would otherwise be removed if we sample the graph using link structure alone.

\section{Conclusion}
\label{sec:conclusion}
We have presented an efficient and extremely simple algorithm for community identification in large-scale graphs by fusing content and link similarity.
Our algorithm, CODICIL, selectively retains edges of high relevancy within local neighborhoods from the fused graph, and subsequently clusters this backbone graph with any content-agnostic graph clustering algorithm. 
%Our method transforms vertex content information into content edges which it merges with original topology edges 
%CODICIL fuses link and content information to improve community detection in moderate and large web graphs.

Our experiments demonstrate that CODICIL outperforms state-of-the-art methods in clustering quality while running orders of magnitude faster for moderately-sized datasets, and can efficiently handle large graphs with millions of nodes and hundreds of millions of edges.
%It can handle topology alone particular this approach is very effective when combined with content
%While the simplification process can be applied 
While simplification can be applied to the original topology alone with a small loss of clustering quality, it is particularly potent when combined with content edges, delivering superior clustering quality with excellent runtime performance.
%We compare CODICIL with two state-of-the-art content-enhanced community detection algorithms which it matches or dominates in clustering quallity while running orders of magnitude faster, for datesets of moderate size.
%It also scales up to large graphs with millions of nodes and hundreds of edges, which current methods do not handle at all.
%Compared to state-of-the-art content-enhanced community detection algorithms
%For moderate web graphs our methods match or dominate state-of-the-art methods while running orders of magnitude faster, and for large web graphs 
%Our methods match or dominate state-of-the-art methods in clustering quality while running orders of magnitude faster on moderate-sized web graphs, and are efficient and scalable enough to handle graphs millions of node, and hundreds of millions of edge graphs which other methods cannot due to memory requirements.

\section{Acknowledgements}
This work is sponsored by NSF SoCS Award \#IIS-1111118, ``Social Media Enhanced Organizational Sensemaking in Emergency Response''.
\bibliographystyle{abbrv}
\bibliography{content_net_spars}
%\balancecolumns

\end{document}